\documentclass[%
 reprint,
amsmath,amssymb,
pra,
]{revtex4-2}

\usepackage{graphicx} 
\usepackage{dcolumn} 
\usepackage{bm} 
\usepackage{hyperref} 
\usepackage{color}
\usepackage{soul}
\usepackage[dvipsnames]{xcolor}
\usepackage{physics}
\graphicspath{ {./} }

\newcommand{\red}[1]{{\color{black} #1}}

\begin{document}

\title{Optomechanical interface between telecom photons and spin quantum memory}

\author{Prasoon K.\ Shandilya}
\thanks{These authors contributed equally}

\author{David P.\ Lake}
\thanks{These authors contributed equally}

\author{Matthew  Mitchell}
\author{Denis D.\ Sukachev}
\author{Paul E.\ Barclay}

\email[Paul~E.\ Barclay: ]{Corresponding author pbarclay@ucalgary.ca}
\affiliation{
Institute for Quantum Science and Technology, University of Calgary, Calgary, Alberta T2N 1N4, Canada}

\date{\today}

\begin{abstract}

Quantum networks enable a broad range of practical and fundamental applications spanning  distributed quantum computing to sensing and metrology. 
A cornerstone of such networks is an interface between telecom photons and quantum memories.
Here we demonstrate a novel approach based on cavity optomechanics that utilizes the susceptibility of spin qubits to strain.
We use it to control electronic spins of nitrogen-vacancy centers in diamond with photons in the 1550\,nm telecommunications wavelength band.
This method does not involve qubit optical transitions and is insensitive to spectral diffusion.
Furthermore, our approach can be applied to solid-state qubits in a wide variety of materials, expanding the toolbox for quantum information processing.
\end{abstract}

\maketitle

\textbf{
Quantum technologies are evolving rapidly, driven by applications in quantum sensing~\cite{Degen-QuantumSensing-RMP-2017}, communications~\cite{Gisin-RMD-2002-QuantumCryptogrpahy}, computing \cite{Ladd2010b}, 
and  networking~\cite{Kimble-Nature-2008-QuantumInternet}. Optically active defects in solids---colour centres---are a promising platform for implementing quantum technologies \cite{Awschalom2018}. Their spin degrees of freedom serve as quantum memories that in some cases can operate at room temperature. When entangled with photons, they form quantum nodes---building blocks of a quantum network \cite{Hensen-Nature-2015-LoopHoleFreeBell1p3km}. This has been achieved with microwave spin control and resonant optical excitation  \cite{Togan-SpinPhotonEntanglement-2010}, but is hindered by broadening of optical transitions from thermal phonons and spectral diffusion~\cite{Faraon2012, Ruf-FPCavity-arxiv-2020}. Furthermore, spin-qubit optical transitions are often outside telecommunications wavelength bands used for long-distance fiber optic transmission. Harnessing coupling between mechanical motion and spins has emerged as an alternative route for controlling spin-qubits~\cite{wang-SpinMechanicsCoupling-2020-arxiv, Lee-SpinMechanicsReview-JoO-2017}. However, connecting spin-mechanical systems to optical links has remained a challenge. Here we use a cavity-optomechanical device \cite{Kippenberg-2014-RMP-CavityOptomechanics} to create a spin-photon interface that does not depend on optical transitions and can be applied to a wide range of spin qubits.
}

Acoustic waves in solids play a key role in practical devices such as modulators, electronic filters, and sensors \cite{Delsing-SAWroadmap-JPhD-2019}. Mechanical degrees of freedom are also central to many quantum technologies, thanks to their ability to couple to a wide range of fields---electrical, magnetic, electromagnetic, and gravitational---through device engineering. For example, phonons mediate quantum gates between trapped ions in quantum computers \cite{Leibfried-IonReview-RevModPhys-2003} and can coherently connect superconducting qubits \cite{Bienfait-PhononSCQubit-Science-2019}. Experiments with spin-qubits have demonstrated that acoustic waves generated piezoelectrically can control electronic spins of diamond and silicon-carbide colour centres in bulk \cite{MacQuarrie-PRL-2013-StrainNVControl, Golter-PRX-2016-NVcoupledSAW, Whiteley-SiCspin-phononCoupling-NatPhys-2019, Maity-2020-NatComm-CoherentContolSiVacoustic}, cantilever  \cite{Mesala-2016-PRAppl-NVmechanics, Ovartchaiyapong2014, Barfuss-StrongMechanicalDrivinfNV-NatPhys-2015}, and hybrid nanowire \cite{Arcizet-NVNanowire-2011-NatPhys, Pigeau-NVcouledToNanowireSpinProtection-NatCom-2015} mechanical resonators, as well as erbium ions in cantilevers~\cite{ Ohta-OptomechanicsREI-PRL-2021}. Despite these advances in spin-mechanical devices, combining them with an interface for optically controlling the mechanical resonator has yet to be realized.  This capability would enable optomechanical spin control, but is challenging due to weak interactions between mechanical resonators and photons.

\begin{figure}[t]
	\includegraphics[width=\linewidth]{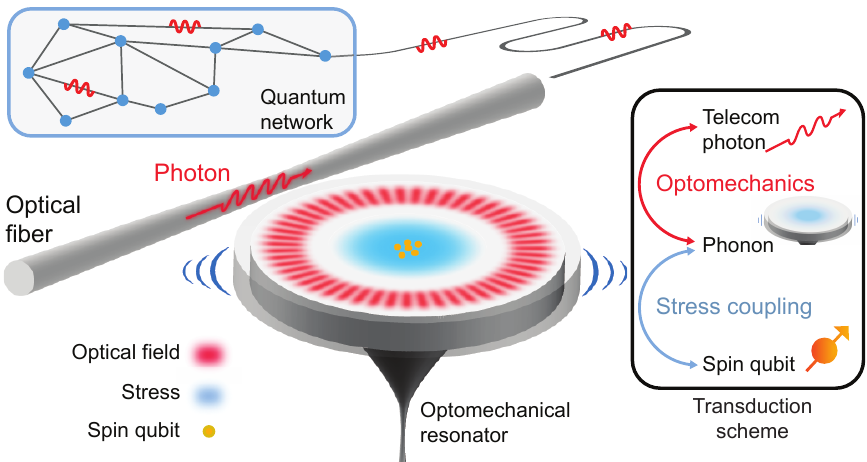}
		\caption{
		    \label{fig1}
		    {\bf The microdisk cavity optomechanical interface between spin quantum memories and telecommunication wavelength photons.} Insets show the transduction scheme (right) and a conceptual large-scale quantum network (left).
	    }
\end{figure}

Cavity optomechanical devices \cite{Kippenberg-2014-RMP-CavityOptomechanics} solve this challenge: by integrating  mechanical resonators within an optical cavity they increase the photon-phonon interaction time and the optomechanical coupling rate ($g_\text{om}$). 
They offer a parametric enhancement of $g_\text{om}$ by increasing the number of intracavity photons $N$, creating a coherent optomechanical interface characterized by optomechanical cooperativity $C_\text{om} = 4 N g_\text{om}^2/\kappa \gamma_\text{m} > 1$, where $\kappa$ and $\gamma_\text{m}$ are the optical cavity and mechanical resonator dissipation rates, respectively.
This regime has been realized in a variety of  cavity optomechanical devices, including those fabricated from diamond~\cite{Burek-DiamondOptomechanicalCrystal-Optica-2016, Mitchell-DiamondOptomechanicalResonator-Optica-2016}.
When cooled near their mechanical ground state, devices with $C_\text{om} > 1$ can generate entanglement between photons and phonons \cite{Cohen-PhononCounting-Nature-2015, Wallucks-QuantumMemoryTelecomBand-NatPhys-2020}.
These devices are promising for universal quantum transducers, for example between optical photons and superconducting microwave resonators \cite{Regal-QuantumTransducer-JoP-2011, Mirhosseini-arxiv-2020-QuantumTransduction, Forsh-MwtoOpticsTransduvtionOptomechanics-NatPhys-2020, Lauk-QuantumTransduction-QScTech-2020}.

In this Article, we couple phonons to both light and electronic spins, creating a cavity optomechanical interface with spin qubits. Using telecommunication wavelength photons and operating at room temperature, we manipulate an ensemble of spin qubits in a diamond microdisk cavity.
This device and the spin-optomechanical interface's operating principle is shown in Fig.\ 1.
Radiation pressure from photons in a microdisk whispering gallery mode coherently excites vibrations of a mechanical mode. 
This motion, approximately described by oscillation of the microdisk diameter, creates a microscopic stress field at the mechanical resonance frequency that interacts with spin qubits in the diamond.
The optomechanical interaction can be tuned for reversible photon-phonon conversion, and can operate at any wavelength resonant with the cavity.
The resulting photon-spin interface does not rely on qubit optical transitions, and we show that it allows manipulation of diamond nitrogen-vacancy (NV) spins \cite{Doherty2013} with telecommunication light. 
Moreover, it can be applied to other color centers, including optically inactive qubits in solids \cite{Soykal-QubitsSiliconPhonons-PRL-2011}, and to manipulate quantum-dot single-photon sources \cite{Yeo--QD-phonon-coupling--NNano-2014}.

\begin{figure}[t!]
	\includegraphics[width=\linewidth]{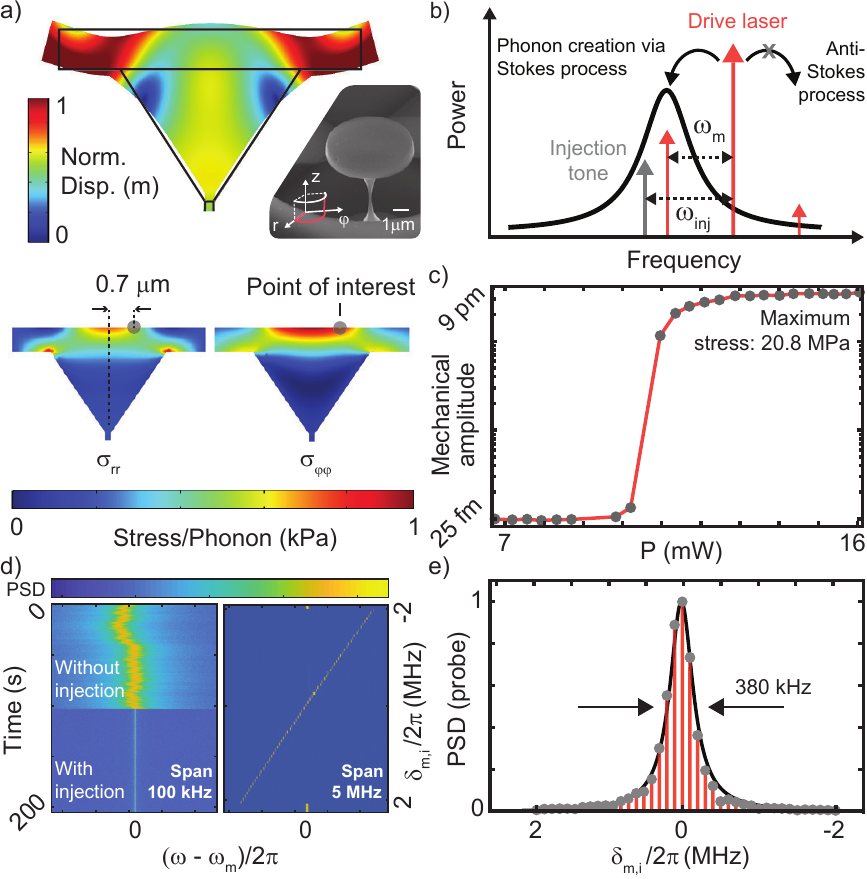}
		\caption{
		    \label{fig2}
		    {\bf Optomechanical characterization of the microdisk resonator.}
		     a) The displacement profile of the RBM and cross-sections for its radial $\sigma_\text{rr}$ and azimuthal $\sigma_{\phi \phi}$ stress tensor components produced by a single phonon. `Point of interest' is the location of the confocal spot for the collection of NV  photoluminescence. The inset shows a scanning electron micrograph image of the microdisk under study.
		     b) Relative frequencies of the relevant optical fields and optical cavity mode (black Lorentzian) during phonon lasing with injection locking.
		     c) Measured amplitude of the optomechanically amplified RBM for varying optical power input to the fiber taper waveguide.
 		     d) Power spectral density (PSD) spectrographs of the RBM displacement for free-running (top-left) and injection-locked (bottom-left) self-oscillations.
 		     Right panel shows the PSD for varying injection locking frequency. 
 		     The color bar spans 90 dBm. 
		     e) Normalized PSD as a function of detuning between the \red{mechanical resonator and the injection locking source} $\delta_\text{m,i} \red{(= \omega_\text{m}-\omega_\text{inj})}$.  The red lines are the PSD spectra for the injection locked RBM at discretely varying $\delta_\text{m,i}$. Grey dots are the peak values of each PSD, which are fit to a Lorentzian.
	}
\end{figure}

\section*{Experiment}

\subsection*{Tunable optomechanical phonon excitation}

Nanophotonic  devices  such  as  microdisks  are naturally suited for spin-optomechanical interfaces. 
Their small size supports GHz frequency mechanical modes that are resonant with a variety of qubits, can be cryogenically cooled to low phonon occupation  \cite{Cohen-PhononCounting-Nature-2015}, and whose low mechanical and optical mode volumes enhance spin-phonon \cite{Mesala-2016-PRAppl-NVmechanics} and photon-phonon coupling rates, respectively.
The 5.3\,$\mu$m diameter microdisk used here minimizes mechanical mode volume while maintaining low optical loss needed for coherent optomechanics. It was patterned from a diamond chip (Element Six, optical grade) using quasi-isotropic plasma etching~\cite{Mitchell-2019-APLphotonics-DiamondMicrodisks}.  An optical mode at wavelength $\lambda_\text{o} = 1564$\,nm with quality factor $Q_\text{o} = 1.1\times10^5$ is used to measure and drive the device's mechanical radial breathing mode (RBM), whose displacement and stress distributions are shown in Fig.\ 2a. From thermomechanical spectroscopy (Supplementary Section II), its frequency and quality factor are  $\omega_\text{m}/2\pi = 2.09$\,GHz and $Q_\text{m} = 4300$, respectively. The optomechanical interaction between these modes has a phonon-photon coupling rate $g_\text{om}/2\pi\sim 25$\,kHz \cite{Mitchell-2019-APLphotonics-DiamondMicrodisks, Mitchell-DiamondOptomechanicalResonator-Optica-2016}. Other optomechanical parameters are summarized in Table S1 (see Supplementary Section II).
The RBM  creates stress predominantly  along the microdisk's radial, $\hat{r}$, and tangential, 
$\hat{\phi}$, unit vectors. Other stress tensor components are an order of magnitude smaller (Supplementary Section IV). This stress field is concentrated at the microdisk centre, where a single phonon is predicted to produce $p_0 \sim 1$\,kPa of stress.
While this is near the state-of-the-art \cite{Lee-SpinMechanicsReview-JoO-2017}, it is too weak for single-phonon driving of NV ground-state spin qubits used here.
Instead, we generate a large coherent phononic state using phonon lasing \cite{Rokhsari-2005-OpticsExpress-SelfOscillations}.

Phonon lasing requires driving a sideband-resolved cavity ($\omega_\text{m} > \kappa$) with a laser blue-detuned $\omega_\text{m}$ from resonance (Fig.\ 2b).
Stokes scattering creates a photon in the optical cavity mode and a phonon in the mechanical mode, while the non-resonant anti-Stokes process is suppressed.
When the drive field is strong enough for $C_\text{om} \ge 1$, the phonon generation rate exceeds its intrinsic dissipation rate $\gamma_\text{m}$, generating mechanical self-oscillations.
Figure\ 2c plots the measured RBM amplitude for varying blue-detuned drive laser power input to the fiber \red{(Supplementary Section IV)}, showing a self-oscillation threshold near 10.2\,mW ($\sim 0.5$\,mW dropped into the cavity).
The maximum displacement of 9\,pm, measured in Fig.\ 2c by calibrating the self-oscillation signal with the thermomechanical signal at low laser power, corresponds to stress $p_\text{max} = 20.8$\,MPa that is large enough to drive diamond NV spin qubits (Supplementary Section IV). 

Efficient spin-optomechanical coupling also requires that mechanical oscillations are resonant with the desired electronic spin transition frequency ($\omega_\text{s}$). 
We coarsely tune $\omega_\text{s}$ with a magnetic field, and then fine-tune the mechanical oscillation frequency using injection locking \cite{Hossein-Zadeh-2008-APL-InjectionLockingOptomechanics} (Supplementary Section II). 
Phase modulating the drive laser at frequency $\omega_\text{inj}$ modulates the intracavity radiation pressure and entrains the self-oscillations. 
Without injection locking, the self-oscillation frequency slowly drifts due to laser and environmental fluctuations  (Fig.\ 2d top-left). With injection locking, the self-oscillation frequency is stable (Fig.\ 2d bottom-left).
Importantly, we can tune the self-oscillation frequency by $\pm 5$ MHz by adjusting the detuning $\delta_\text{m,i} = \omega_\text{m}-\omega_\text{inj}$ between the intrinsic mechanical frequency and the injection tone (Fig.\ 2d right).
However, as shown in Fig.\ 2e, the oscillations' power spectral density decreases with increasing $\abs{\delta_\text{m,i}}$ ~\cite{Hong-InjectionLocking-IEEE-2019}, following a Lorentzian profile with FWHM $2\pi \times 380$\,kHz close to the mechanical resonance's intrinsic bandwidth $\gamma_\text{m} = \omega_\text{m} / Q_\text{m} \approx 2\pi \times 490$\,kHz. 
Since the largest self-oscillation amplitude occurs when the injected tone and mechanical frequency are resonant ($\delta_\text{m,i} = 0$), to maximize optomechanical spin-driving, the spin-mechanics detuning $\delta_\text{s,m} = \omega_\text{s} - \omega_\text{m}$ should be zeroed via the magnetic field.
\red{The combination of optomechanical direct driving and cavity backaction used to control the mechanical resonator
shares advantages of both: the tunability of direct driving and the relative simplicity of exciting stable large-amplitude self-oscillations.}

\begin{figure}[t]
	\includegraphics[width=\linewidth]{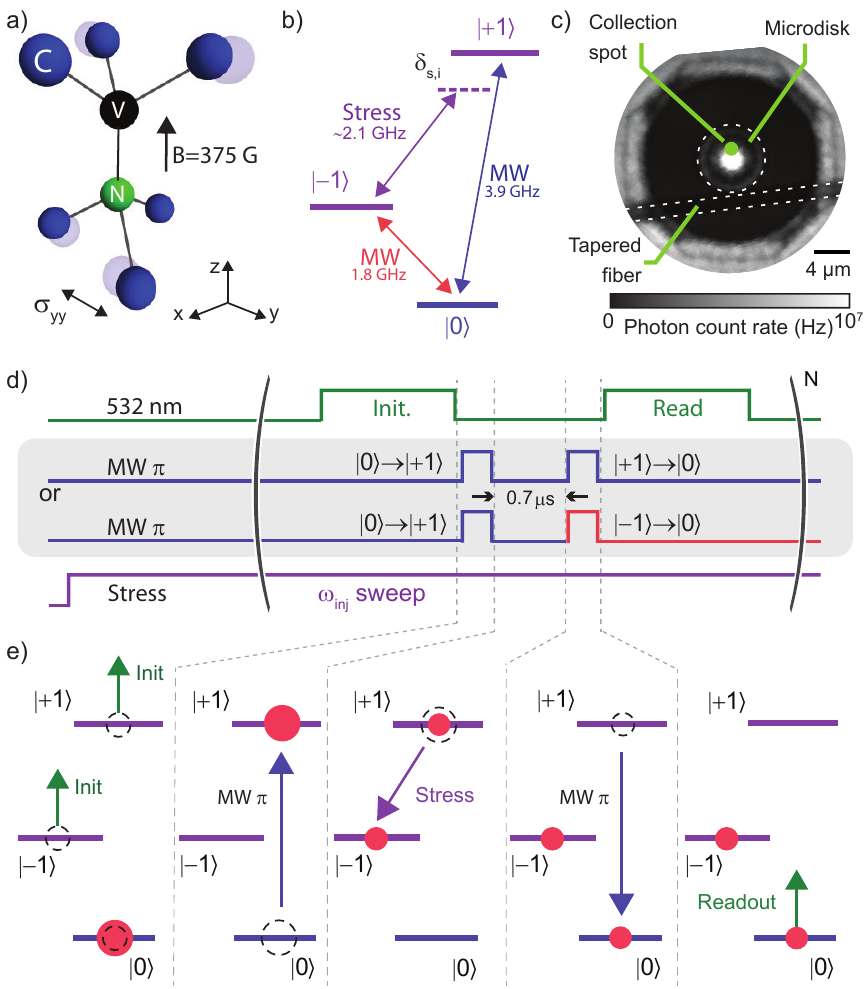}
		\caption{
		    \label{fig3}
		    {\bf Nitrogen-vacancy centers in diamond and spin driving sequence.}
		    a) Atomic structure of the NV center. Grey circles represent the displacement of carbon atoms under a stress along the $y$-axis ($\sigma_\text{yy}$). 
		    b) The energy level diagram of the NV ground state spin triplet at a magnetic field $B=375$\,G applied along the NV symmetry axis ($z$); $\delta_\text{s,i}$ is the detuning between the injection locked mechanical self-oscillation frequency and the $\ket{+1} \rightarrow \ket{-1}$ spin transition.
		    c) A photoluminescence scan over the microdisk, showing the position of the NV measurement spot.
		    d) Pulse sequence used during measurements of optomechanical NV spins driving.
		    e) Population of NV levels at each step of the pulse sequence in d). The diameter of each solid red (dashed black) circle represents population after (before) each step.
		    }
\end{figure}


\subsection*{Spin-optomechanical driving}

We next use this tunable phonon lasing to demonstrate a spin-optomechanical interface.
The negatively charged NV center has an electronic spin-triplet ground state $\{ \ket{0}, \ket{\pm1} \}$  (Fig.\ 3a) that can be optically initialized and \red{read out} at room temperature~\cite{Gruber-Science-1997-SingleNVdetection}.
Mechanical coupling to NV spins arises from  deformation of their molecular orbitals by crystal lattice strain that displaces carbon atoms surrounding the NV defect (Fig.\ 3b)~\cite{Udvarhelyi-PRB-2018-NVspinStrainInteractions, wang-SpinMechanicsCoupling-2020-arxiv}.
Coupling NV centers to mechanical resonators \cite{MacQuarrie-PRL-2013-StrainNVControl, MacQuarrie-2015-Optica-CoherentControlNVstrain, Golter-PRX-2016-NVcoupledSAW, Ovartchaiyapong2014, Hong-2012-NanoLetters-NVcoupledCanteliver}  has led to new spin manipulation capabilities~\cite{MacQuarrie-2015-PRB-DynamicalDecouplingSpin, MacQuarrie-2018-PRL-NVorbitalStrainManipulation},  techniques for suppressing decoherence \cite{Pigeau-NVcouledToNanowireSpinProtection-NatCom-2015, Barfuss-StrongMechanicalDrivinfNV-NatPhys-2015}, and tuning of NV emission \cite{Lee-NVcanteliverTuning-PRApplied-2016}. 
However, the lack of optical cavities in these systems has prevented interfacing them with photons.
Diamond microdisks provide this needed element, enabling optomechanical manipulation of the strain-coupled \red{$\ket{+1} \rightarrow \ket{-1}$ transition}. 
\red{Each of these levels is split by interaction with the $^{14}$N nuclear spin (see Supplementary Fig.\,S5). Since strain coupling preserves nuclear spin, we focus on one of these transitions~\cite{MacQuarrie-PRL-2013-StrainNVControl}, choosing the zero nuclear spin hyperfine level to simplify initialization and read out (Supplementary Section III). Note that the other transitions are $\pm 4$ MHz detuned, where the injection locked mechanical self-oscillation amplitude is too small to efficiently drive them.
}

To study spin-optomechanical coupling, the device was mounted in a confocal microscope (0.8 NA) operating in ambient conditions.
A 532 nm laser is used to for NV initialization and read out. Figure 3c shows an image obtained from photoluminescence upon rastering the laser over the microdisk.
A 375\,G magnetic field from a permanent magnet, aligned along one of four possible NV crystallographic orientations, splits the $\ket{\pm 1}$ levels of this subset of NVs close to resonance with $\omega_\text{m}$. 
A thin wire delivers microwave pulses for spin measurement sequences discussed below (Supplementary Section III).

For NVs at the point of maximum stress at the microdisk center (Fig.\ 2a), we predict that mechanical self-oscillations drive the spin transition at rate  $ \Omega_\text{m}^\text{max} \approx 2\pi \times p_0 g_\text{str} \approx 2\pi \times 395$\,kHz, where  $g_\text{str} \approx 19$\,Hz/kPa is the NV-stress susceptibility \cite{Ovartchaiyapong2014}. 
Unfortunately, photoluminescence at the device center is dominated by NVs in the pedestal, as shown in Fig.\ 3c. %
These NVs are uncoupled to the RBM, and their emission degrades the signal from microdisk NVs.
As a compromise, we study NVs offset 0.7\,$\mu$m from the microdisk center, as indicated in Figs.\ 1c and 3c. Here stress is reduced by  $<30\%$, and pedestal photoluminescence is sufficiently suppressed to observe spin-mechanical coupling.
\red{Taking into account the direction of the NV symmetry axis and the tensorial {nature of} NV-stress interaction, we estimate $\Omega_\text{m} / 2\pi \approx 100$\,kHz at this location (Supplementary Section IV).}

\begin{figure}[t!]
	\includegraphics[width=\linewidth]{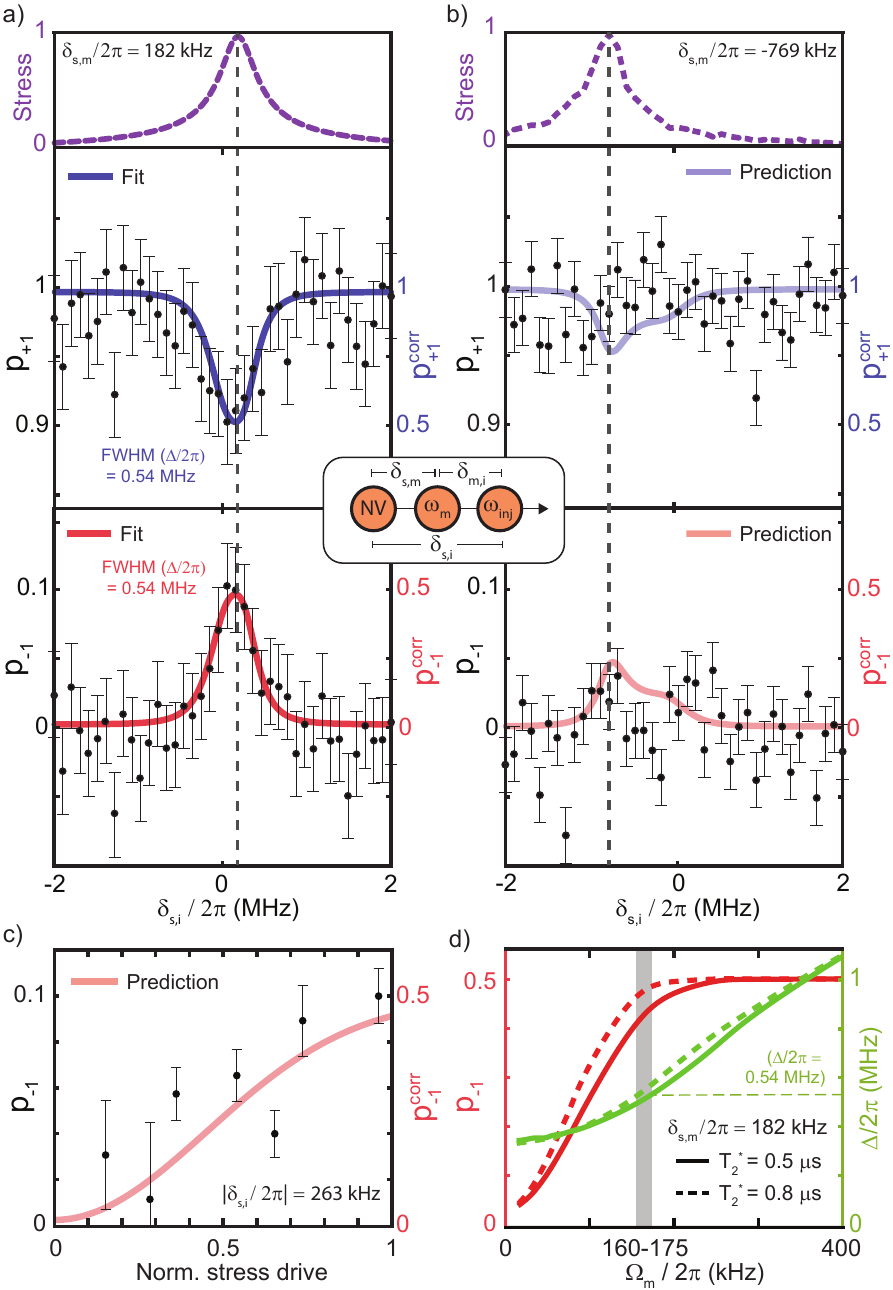}
		\caption{
		    \label{fig4}
		    {\bf Optomechanical control of NV centers.} 
		    a) Measured populations of $\ket{\pm 1}$ when the self-oscillation injection locking frequency and hence $\delta_\text{s,i}$ is varied, for 7\,$\mu$s of mechanical drive and $\delta_\text{s,m} / 2 \pi = 182$\,kHz. The top panel shows the corresponding measured variation in optomechanically induced stress.
		    Solid lines are fits to the model \red{with $r$ being a free parameter and $T_2^* = 0.8\,\mu$s} (Supplementary Sections IV and V). 
		    b) Same as a) for $\delta_\text{s,m} / 2\pi = - 769$\,kHz. Solid lines are theoretical predictions based on parameters from a).
		    The inset shows the detunings between the NV spin transition, the intrinsic mechanical frequency, and the injection locking tone.
		    \red{Dashed vertical lines in a) and b) denote detunings where mechanical stress is maximum ($\delta_\text{m,i} = 0$).}
		    c) Measured population of $\ket{- 1}$ for varying optomechanically induced stress after 7\,$\red{\mu}$s of mechanical drive when $\abs{\delta_\text{s,i}} / 2\pi = 263$\,kHz.
		    d) Simulated FWHM ($\Delta$) and population change for different spin-stress coupling strength $\Omega_\text{m}$, generated using the $\delta_\text{s,i}$ dependent stress amplitude from (a) for $T_2^* =  0.5$, $0.8$ $\mu$s,  $\delta_\text{s,m} / 2\pi = 182$\,kHz, \red{ and $r = 0$}.
		    \red{The horizontal dashed line denotes $\Delta / 2\pi = 0.54$\,MHz; the vertical gray region represents the estimated range of $\Omega_m$.}
		    Error bars are one standard deviation.
	    }
\end{figure}

To measure stress-induced driving of NV spins, we apply the pulse sequence in Fig.\ 3d. 
First, NVs are pumped into $\ket{0}$ with the 532 nm laser (Fig.\ 3e). 
Then a microwave $\pi$-pulse transfers spins from $\ket{0}$ to $\ket{+1}$. 
We then mechanically drive $\ket{+1} \rightarrow \ket{-1}$ for 7 $\mu$s at a frequency set by the injection locking tone $\omega_\text{inj}$. 
Finally, we measure the population $p_{+1}$ remaining in $\ket{+1}$ by transferring it back to $\ket{0}$ with a second $\pi$-pulse, followed by reading out $\ket{0}$ using a green laser pulse. 
During this sequence, the mechanical driving depletes the population in $\ket{+1}$ by promoting it to $\ket{-1}$.
Upon read out, this appears as missing population in $\ket{0}$. \red{Note that the $\omega_\text{inj}$ tone and mechanical drive amplitude are constant during this sequence.}
The scheme is then repeated with the microwave $\pi$-pulses modified to \red{read out} the population $p_{-1}$ in $\ket{-1}$ mechanically transferred from spins initialized in $\ket{+1}$.

As described above, the mechanical amplitude is maximum when $\delta_\text{m,i} = 0$ (Fig.\ 2e).
However, efficient spin-stress driving requires the spin transition be resonant with the injection locked oscillations: $\delta_\text{s,i} = 0$.
Both conditions are met if $\delta_\text{s,m} = 0$ through magnetic field tuning. In our apparatus the closest to this condition that we achieved \red{during a measurement run} was $\delta_\text{s,m} = 2\pi \times 182$\,kHz \red{due to imprecision in positioning the magnet}.
We then measured $p_{\pm 1}$ as a function of $\delta_\text{s,i}$ by varying $\omega_\text{inj}$ (Fig.\ 4a).
The coinciding dip in $p_{+1}$ and peak in  $p_{-1}$, together with their dependence on $\delta_\text{s,i}$, verifies that spins are being optomechanically driven.
As a control dataset, we repeated the same measurements for $\omega_\text{m}$ far from the spin resonance, $\delta_\text{s,m} / 2\pi =  -769$\,kHz (Fig.\ 4b).
Setting the injection locking detuning to compensate for the spin-mechanics detuning, $\delta_\text{m,i} = - \delta_\text{m,s}$, brings the mechanical oscillation frequency back to spin resonance, but at a significantly reduced amplitude, as shown in Fig.\ 2e.
Within the measurement's signal-to-noise ratio, we were not able to reliably identify any peak or dip in this case, supporting the above claims.
Next, we repeated these measurements for varying mechanical amplitude by changing $\delta_\text{s,m}$ while keeping $ \delta_\text{s,i}$ constant via adjustment of $\omega_\text{inj}$ (Fig.\ 4c). This was possible thanks to slow drift in $\delta_\text{s,m}$ over the course of several measurement runs (Supplementary Section V).
As expected, the transferred spin populations monotonically increase with stress amplitude.

\subsection*{Estimating the spin-mechanics coupling rate}

To extract the spin-mechanics coupling rate $\Omega_\text{m}$ from the measurements, \red{we compare them with predictions from a quantum master equation model  (Supplementary Sections IV and V). This model, which includes the spin transition dephasing ({$1/\pi T_2^* = 400\,\text{kHz}$}) and  the injection locking bandwidth $\Gamma_\text{tune}$, generates the theoretical curves in Fig.\ 4. Its fitting parameters are $\Omega_\text{m}$ and the fraction $r$ of photoluminescence from pedestal NVs uncoupled to the mechanical resonance ({Fig.\,3c}, Supplementary Section V).
In our parameter regime, $\Omega_\text{m} < 2\pi/T_2^*$, the width (FWHM $\Delta$) of the dip (peak) in $p_{+1}$ ($p_{-1}$) is determined predominantly by $\Gamma_\text{tune}$ and $\Omega_\text{m}$.
Figure 4d shows the predicted dependence of $\Delta$ on $\Omega_\text{m}$. Given $\Delta /2\pi \approx 540~\text{kHz}$ observed in Fig.\ 4a,  {we infer} $\Omega_\text{m} / 2\pi = 168\pm8~\text{kHz}$ 
(error due to uncertainty in $T_2^*$, 
discussed in Supplementary  Section  III). }
\red{For this coupling rate, Fig.\ 4d also shows that  a $45$\% change in $p_{-1}$ is expected. This is significantly larger than the  $10\%$ change observed in Fig.\ 4a, and can be explained by pedestal NVs reducing the measured contrast.
To account for this, we define corrected spin populations $p_{- 1}^\text{corr} = p_{- 1}/(1-r)$ and $p_{+1}^\text{corr} = 1-p_{-1}^\text{corr}$. 
For $p_{-1}^\text{corr}$ to agree with the 45\% prediction, we estimate $r\approx 0.8$. A secondary vertical axis for $p_{\pm 1}^\text{corr}$ obtained for this $r$ is included in Figs.\,4a and 4b.
\red{Note that a conservative lower bound on $\Omega_\text{m}$ can be inferred
by ignoring $\Delta$ and using the uncorrected ($r = 0$) change of $10\%$ in $p_{-1}$, which Fig.\ 4d shows corresponds to $\Omega_\text{m}^\text{min} / 2\pi \approx 50$\,kHz.}

The theoretically predicted $p_{\pm 1}$  for the values of $\Omega_\text{m}$ and $r$ extracted above are plotted in Figs.\ 4a-c, and agree with the measured data within our signal-to-noise. The far-detuned control data in Fig.\ 4b would ideally exhibit a small signal, but it is not resolvable due to noise.
Note that we have assumed that the spins are distinguishable due to inhomogeneous broadening, and as such there is no multispin enhancement due to the spin-phonon coupling in the experiment.}

\red{In future, more precise determination of $\Omega_\text{m}$ can be obtained by measuring single NVs \cite{Ovartchaiyapong2014} in ultrapure diamond devices. This would eliminate the need for $r$ and allow spatial imaging of $\Omega_\text{m}$ \cite{Whiteley-SiCspin-phononCoupling-NatPhys-2019}. 
Measuring $p_{\pm1}$ for varying interaction time would provide further characterization of $\Omega_\text{m}$. However, relatively low signal-to-noise (SNR) in our setup prevents observing smaller signals obtained for shorter interaction times.
For longer interactions, the signal does not change significantly yet noise increases. In all measurements, drift in $\omega_\text{m}$ and $\omega_\text{s}$ prevented boosting SNR via additional data acquisition and averaging. Active temperature and magnetic field tuning could reduce these noise sources. }

\red{Increasing $\Omega_\text{m}$ offers a more favorable approach to boosting the SNR, and would also allow observation of coherent oscillations of spin population \cite{MacQuarrie-2015-Optica-CoherentControlNVstrain}. 
Presently, $\Omega_\text{m}$ is limited by clamping of the self-oscillation amplitude by nonlinearities in the optical mode's optomechanical transduction \cite{Poot-PRA-BackactionLimitSO-2012}.
Direct optomechanical driving with a large amplitude tone will be similarly limited by nonlinear effects.
The nonlinearity can be reduced and the clamped amplitude can be increased using a lower $Q_\text{o}$ mode at the expense of requiring a higher drive laser power.  Alternatively, creating devices incorporating spins with higher strain sensitivity and supporting more localized mechanical modes can increase $\Omega_\text{m}$, as discussed below.}


\section*{Discussion and future directions}

\subsection*{Creating a single photon interface}

The device demonstrated here is a proof-of-principle optomechanical interface between classical light and spin qubits. Many of its quantum networking applications require operation in a regime where a single photon coherently \red{and reversibly} couples to a spin qubit. \red{Reaching this regime requires } $C_\text{om}$ and spin-mechanical cooperativity \red{($C_\text{sm}$)} both $ > 1$ \cite{Lauk-QuantumTransduction-QScTech-2020}. 
\red{
As summarized in Fig.\ 5, reaching this quantum regime is possible using already demonstrated diamond cavity optomechanical devices coupled to diamond spin states with higher stress-sensitivity than the NV ground state.

\begin{figure}[b]
	\includegraphics[width=\linewidth]{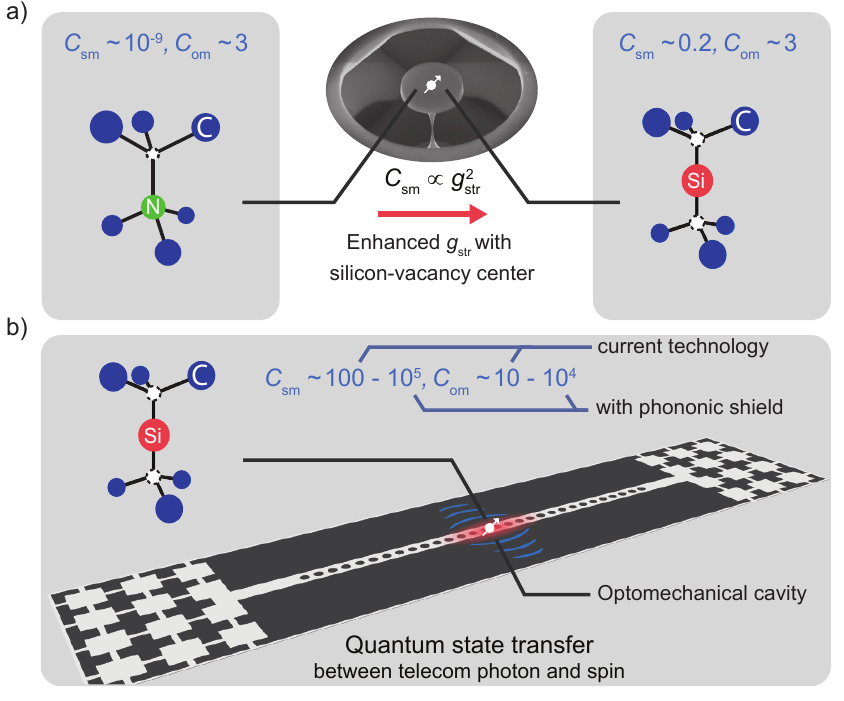}
		\caption{
		    \label{fig5}
		    \red{
		    {\bf Roadmap for realizing a single photon spin-optomechanical interface.} a) Using diamond silicon-vacancy centers as a spin qubit will increase $C_\text{sm}$ above 0.1 thanks to their $\sim 10^5$ times higher strain sensitivity. b) Replacing the microdisk with an optomechanical crystal will further increase $C_\text{sm}$ by more than $\times 10^3$, and a phononic shield will further boost both $C_\text{sm}$ and $C_\text{om}$. Similarly high cooperativity can be realized using NV ground state spins coupled through a phonon assisted optical Raman process. }
	    }
\end{figure}

Optomechanical $C_\text{om} > 1$ is routinely demonstrated in diamond microdisks, as verified by the onset of self-oscillations in our current devices.
Reversible photon-phonon conversion is achieved via optomechanically-induced transparency, where the drive laser is red-detuned (in contrast to blue detuning used above for generating self-oscillations). This regime has already been demonstrated and explored with diamond microdisks~\cite{lake2021processing}.
The outstanding technical challenge is performing coherent conversion in a device cryogenically cooled near its mechanical quantum ground state without heating it via material optical absorption. This has been demonstrated in silicon optomechanical quantum memories \cite{Cohen-PhononCounting-Nature-2015, Wallucks-QuantumMemoryTelecomBand-NatPhys-2020}. In diamond devices it will be aided by the material's low nonlinear absorption and excellent thermal properties. }

\red{
Realizing spin-mechanical $C_\text{sm} = g_\text{sm}^2/\gamma_\text{m}\gamma_\text{spin} > 1$ in our device is hindered by the NV ground state's intrinsically weak spin-phonon coupling, resulting in  $C_\text{sm} \sim 10^{-9}$.
Fortunately, as detailed in Supplementary Section VI and in Fig.\ 5a, realizing $C_\text{sm} > 1$ is possible by coupling phonons to NV excited states, or to diamond silicon-vacancy (SiV) ground states. Both of these systems are approximately $\times 10^5$ more sensitive to stress than the NV ground state. However they require low temperature operation for good coherence, making them less convenient for the proof-of-concept spin-optomechanics experiment presented here.

We focus primarily on the SiV since it has been successfully incorporated into nanophotonic devices while maintaining excellent spin and optical properties. The interaction between SiV spins and GHz frequency phonons has been well characterized \cite{Maity-2020-NatComm-CoherentContolSiVacoustic}.  We predict that a single phonon of the radial breathing mode will couple to an SiV spin with rate $g_\text{sm} / 2\pi \sim 0.1$ MHz, using either a Raman phonon-microwave coupling scheme \cite{neuman-PhotonicBus-2020-arxiv}, or direct coupling of a magnetically tuned qubit \cite{Meesala2018}. This corresponds to $C_\text{sm} \sim 0.2$ for  $\gamma_\text{spin}/2\pi = 1\,\text{MHz}$, typical for SiV spin qubits, and $\gamma_\text{m}/2\pi = 200\,\text{kHz}$, as measured for nominally identical diamond microdisks in  Ref.\ \cite{lake2021processing}.

As shown in Fig.\ 5b, device improvements will further enhance both $C_\text{sm}$ and $C_\text{om}$.
A diamond optomechanical crystal cavity \cite{Burek-DiamondOptomechanicalCrystal-Optica-2016} will increase $C_\text{sm}$ $> 100$  thanks to its smaller mechanical mode volume and correspondingly larger $g_\text{sm}/2\pi \sim 2-3$\,MHz. These devices would similarly increase $g_\text{om} / 2\pi$ to $\sim 200$\,kHz, resulting in $C_\text{om} \sim 10$, as demonstrated in Ref.\ \cite{Burek-DiamondOptomechanicalCrystal-Optica-2016}. 
A phononic shield could dramatically reduce $\gamma_\text{m}$, increasing both $C_\text{sm}$  and $C_\text{om}$ by several orders of magnitude; $Q_\text{m} > 10^9$ has been observed in phononic crystal shielded silicon optomechanical devices \cite{MacCabe-UltraLongPhononicLifetime-Science-2020}, enabling $C_\text{om} > 1$ for single  ($N \le 1$) drive photons.

Comparable $C_\text{sm}$ could be realized with NV ground state spins coupled {to a strain sensitive NV excited state} using an optical Raman  process \cite{Golter-PRX-2016-NVcoupledSAW}.  
However, spectral diffusion characteristic of optical transitions of NVs in nanophotonic devices would pose a challenge \cite{Ruf-FPCavity-arxiv-2020}.
Finally, dynamically modulating the mechanical frequency through the optomechanical spring effect can offer an exponential enhancement to $g_\text{sm}$ \cite{li2020enhancing}, providing a complementary approach to increasing $C_\text{sm}$.
}

\vspace{12 pt}
\subsection*{Conclusion}
The spin-optomechanical interface demonstrated here allows operation at telecommunication wavelengths regardless of the qubit's resonant transitions, offers protection from spectral diffusion~\cite{Ruf-FPCavity-arxiv-2020}, and can be used with qubits lacking optical transitions~\cite{Soykal-QubitsSiliconPhonons-PRL-2011}.
\red{It is an important step towards} achieving single-photon coherent optomechanical coupling to spins, \red{which will lead to universal coupling of telecom photons to hybrid quantum nodes built on spins and superconducting qubits \cite{neuman-PhotonicBus-2020-arxiv}.} It will also enable non-linear optomechanics via spin induced anharmonicity of the mechanical oscillator, in analogy with superconducting qubits in microwave resonators, potentially leading to hybrid acoustic quantum computers~\cite{Chamberland--CatStatesMW--Mechanics-2020-arxiv}. Other applicantions include spin-cooling of mechanical resonators~\cite{MacQuarrie--SpinCoolingOptomechanics-2017-NatComm}, optomechanical control and detection of single photons~\cite{Kettler-OptomechanicalControlQD-NatNano-2020, Ghobadi-NVoptomechanicalInterfcace-PRA-2019} and  phonons~\cite{MacCabe-UltraLongPhononicLifetime-Science-2020}, and creating quantum transducers \cite{Lauk-QuantumTransduction-QScTech-2020}.

\subsection*{Acknowledgements}

This work was supported by the Alberta Innovates Strategic Research Projects program, the National Research Council Nanotechnology Research Centre, and the NSERC Discovery Grant, Accelerator, CREATE, Strategic Partnership Grant, and RTI programs. The authors acknowledge Harishankar Jayakumar, JP Hadden, Tamiko Masuda, and Behzad Khanaliloo for their contributions to the initial setup of the experimental apparatus.
\\
\\
{\bf Data availability}\\
All data related to the current study are available from the corresponding author on reasonable request.


\begin{thebibliography}{57}%
\makeatletter
\providecommand \@ifxundefined [1]{%
 \@ifx{#1\undefined}
}%
\providecommand \@ifnum [1]{%
 \ifnum #1\expandafter \@firstoftwo
 \else \expandafter \@secondoftwo
 \fi
}%
\providecommand \@ifx [1]{%
 \ifx #1\expandafter \@firstoftwo
 \else \expandafter \@secondoftwo
 \fi
}%
\providecommand \natexlab [1]{#1}%
\providecommand \enquote  [1]{``#1''}%
\providecommand \bibnamefont  [1]{#1}%
\providecommand \bibfnamefont [1]{#1}%
\providecommand \citenamefont [1]{#1}%
\providecommand \href@noop [0]{\@secondoftwo}%
\providecommand \href [0]{\begingroup \@sanitize@url \@href}%
\providecommand \@href[1]{\@@startlink{#1}\@@href}%
\providecommand \@@href[1]{\endgroup#1\@@endlink}%
\providecommand \@sanitize@url [0]{\catcode `\\12\catcode `\$12\catcode
  `\&12\catcode `\#12\catcode `\^12\catcode `\_12\catcode `\%12\relax}%
\providecommand \@@startlink[1]{}%
\providecommand \@@endlink[0]{}%
\providecommand \url  [0]{\begingroup\@sanitize@url \@url }%
\providecommand \@url [1]{\endgroup\@href {#1}{\urlprefix }}%
\providecommand \urlprefix  [0]{URL }%
\providecommand \Eprint [0]{\href }%
\providecommand \doibase [0]{https://doi.org/}%
\providecommand \selectlanguage [0]{\@gobble}%
\providecommand \bibinfo  [0]{\@secondoftwo}%
\providecommand \bibfield  [0]{\@secondoftwo}%
\providecommand \translation [1]{[#1]}%
\providecommand \BibitemOpen [0]{}%
\providecommand \bibitemStop [0]{}%
\providecommand \bibitemNoStop [0]{.\EOS\space}%
\providecommand \EOS [0]{\spacefactor3000\relax}%
\providecommand \BibitemShut  [1]{\csname bibitem#1\endcsname}%
\let\auto@bib@innerbib\@empty
\bibitem [{\citenamefont {Degen}\ \emph {et~al.}(2017)\citenamefont {Degen},
  \citenamefont {Reinhard},\ and\ \citenamefont
  {Cappellaro}}]{Degen-QuantumSensing-RMP-2017}%
  \BibitemOpen
  \bibfield  {author} {\bibinfo {author} {\bibfnamefont {C.~L.}\ \bibnamefont
  {Degen}}, \bibinfo {author} {\bibfnamefont {F.}~\bibnamefont {Reinhard}},\
  and\ \bibinfo {author} {\bibfnamefont {P.}~\bibnamefont {Cappellaro}},\
  }\bibfield  {title} {\bibinfo {title} {{Quantum sensing}},\ }\href
  {https://doi.org/10.1103/RevModPhys.89.035002} {\bibfield  {journal}
  {\bibinfo  {journal} {Reviews of Modern Physics}\ }\textbf {\bibinfo {volume}
  {89}},\ \bibinfo {pages} {35002} (\bibinfo {year} {2017})}\BibitemShut
  {NoStop}%
\bibitem [{\citenamefont {Gisin}\ \emph {et~al.}(2002)\citenamefont {Gisin},
  \citenamefont {Ribordy}, \citenamefont {Tittel},\ and\ \citenamefont
  {Zbinden}}]{Gisin-RMD-2002-QuantumCryptogrpahy}%
  \BibitemOpen
  \bibfield  {author} {\bibinfo {author} {\bibfnamefont {N.}~\bibnamefont
  {Gisin}}, \bibinfo {author} {\bibfnamefont {G.}~\bibnamefont {Ribordy}},
  \bibinfo {author} {\bibfnamefont {W.}~\bibnamefont {Tittel}},\ and\ \bibinfo
  {author} {\bibfnamefont {H.}~\bibnamefont {Zbinden}},\ }\bibfield  {title}
  {\bibinfo {title} {{Quantum cryptography}},\ }\href
  {https://doi.org/10.1103/RevModPhys.74.145} {\bibfield  {journal} {\bibinfo
  {journal} {Reviews of Modern Physics}\ }\textbf {\bibinfo {volume} {74}},\
  \bibinfo {pages} {145} (\bibinfo {year} {2002})},\ \Eprint
  {https://arxiv.org/abs/0101098} {arXiv:0101098 [quant-ph]} \BibitemShut
  {NoStop}%
\bibitem [{\citenamefont {Ladd}\ \emph {et~al.}(2010)\citenamefont {Ladd},
  \citenamefont {Jelezko}, \citenamefont {Laflamme}, \citenamefont {Nakamura},
  \citenamefont {Monroe},\ and\ \citenamefont {O'Brien}}]{Ladd2010b}%
  \BibitemOpen
  \bibfield  {author} {\bibinfo {author} {\bibfnamefont {T.~D.}\ \bibnamefont
  {Ladd}}, \bibinfo {author} {\bibfnamefont {F.}~\bibnamefont {Jelezko}},
  \bibinfo {author} {\bibfnamefont {R.}~\bibnamefont {Laflamme}}, \bibinfo
  {author} {\bibfnamefont {Y.}~\bibnamefont {Nakamura}}, \bibinfo {author}
  {\bibfnamefont {C.}~\bibnamefont {Monroe}},\ and\ \bibinfo {author}
  {\bibfnamefont {J.~L.}\ \bibnamefont {O'Brien}},\ }\bibfield  {title}
  {\bibinfo {title} {{Quantum computers}},\ }\href
  {https://doi.org/10.1038/nature08812} {\bibfield  {journal} {\bibinfo
  {journal} {Nature}\ }\textbf {\bibinfo {volume} {464}},\ \bibinfo {pages}
  {45} (\bibinfo {year} {2010})}\BibitemShut {NoStop}%
\bibitem [{\citenamefont {Kimble}(2008)}]{Kimble-Nature-2008-QuantumInternet}%
  \BibitemOpen
  \bibfield  {author} {\bibinfo {author} {\bibfnamefont {H.~J.}\ \bibnamefont
  {Kimble}},\ }\bibfield  {title} {\bibinfo {title} {{The quantum internet}},\
  }\href {https://doi.org/10.1038/nature07127} {\bibfield  {journal} {\bibinfo
  {journal} {Nature}\ }\textbf {\bibinfo {volume} {453}},\ \bibinfo {pages}
  {1023} (\bibinfo {year} {2008})},\ \Eprint {https://arxiv.org/abs/0806.4195}
  {arXiv:0806.4195} \BibitemShut {NoStop}%
\bibitem [{\citenamefont {Awschalom}\ \emph {et~al.}(2018)\citenamefont
  {Awschalom}, \citenamefont {Hanson}, \citenamefont {Wrachtrup},\ and\
  \citenamefont {Zhou}}]{Awschalom2018}%
  \BibitemOpen
  \bibfield  {author} {\bibinfo {author} {\bibfnamefont {D.~D.}\ \bibnamefont
  {Awschalom}}, \bibinfo {author} {\bibfnamefont {R.}~\bibnamefont {Hanson}},
  \bibinfo {author} {\bibfnamefont {J.}~\bibnamefont {Wrachtrup}},\ and\
  \bibinfo {author} {\bibfnamefont {B.~B.}\ \bibnamefont {Zhou}},\ }\bibfield
  {title} {\bibinfo {title} {{Quantum technologies with optically interfaced
  solid-state spins}},\ }\href {https://doi.org/10.1038/s41566-018-0232-2}
  {\bibfield  {journal} {\bibinfo  {journal} {Nature Photonics}\ }\textbf
  {\bibinfo {volume} {12}},\ \bibinfo {pages} {516} (\bibinfo {year}
  {2018})}\BibitemShut {NoStop}%
\bibitem [{\citenamefont {Hensen}\ \emph {et~al.}(2015)\citenamefont {Hensen},
  \citenamefont {Bernien}, \citenamefont {Drea{\'{u}}}, \citenamefont
  {Reiserer}, \citenamefont {Kalb}, \citenamefont {Blok}, \citenamefont
  {Ruitenberg}, \citenamefont {Vermeulen}, \citenamefont {Schouten},
  \citenamefont {Abell{\'{a}}n}, \citenamefont {Amaya}, \citenamefont
  {Pruneri}, \citenamefont {Mitchell}, \citenamefont {Markham}, \citenamefont
  {Twitchen}, \citenamefont {Elkouss}, \citenamefont {Wehner}, \citenamefont
  {Taminiau},\ and\ \citenamefont
  {Hanson}}]{Hensen-Nature-2015-LoopHoleFreeBell1p3km}%
  \BibitemOpen
  \bibfield  {author} {\bibinfo {author} {\bibfnamefont {B.}~\bibnamefont
  {Hensen}}, \bibinfo {author} {\bibfnamefont {H.}~\bibnamefont {Bernien}},
  \bibinfo {author} {\bibfnamefont {A.~E.}\ \bibnamefont {Drea{\'{u}}}},
  \bibinfo {author} {\bibfnamefont {A.}~\bibnamefont {Reiserer}}, \bibinfo
  {author} {\bibfnamefont {N.}~\bibnamefont {Kalb}}, \bibinfo {author}
  {\bibfnamefont {M.~S.}\ \bibnamefont {Blok}}, \bibinfo {author}
  {\bibfnamefont {J.}~\bibnamefont {Ruitenberg}}, \bibinfo {author}
  {\bibfnamefont {R.~F.}\ \bibnamefont {Vermeulen}}, \bibinfo {author}
  {\bibfnamefont {R.~N.}\ \bibnamefont {Schouten}}, \bibinfo {author}
  {\bibfnamefont {C.}~\bibnamefont {Abell{\'{a}}n}}, \bibinfo {author}
  {\bibfnamefont {W.}~\bibnamefont {Amaya}}, \bibinfo {author} {\bibfnamefont
  {V.}~\bibnamefont {Pruneri}}, \bibinfo {author} {\bibfnamefont {M.~W.}\
  \bibnamefont {Mitchell}}, \bibinfo {author} {\bibfnamefont {M.}~\bibnamefont
  {Markham}}, \bibinfo {author} {\bibfnamefont {D.~J.}\ \bibnamefont
  {Twitchen}}, \bibinfo {author} {\bibfnamefont {D.}~\bibnamefont {Elkouss}},
  \bibinfo {author} {\bibfnamefont {S.}~\bibnamefont {Wehner}}, \bibinfo
  {author} {\bibfnamefont {T.~H.}\ \bibnamefont {Taminiau}},\ and\ \bibinfo
  {author} {\bibfnamefont {R.}~\bibnamefont {Hanson}},\ }\bibfield  {title}
  {\bibinfo {title} {{Loophole-free Bell inequality violation using electron
  spins separated by 1.3 kilometres}},\ }\href
  {https://doi.org/10.1038/nature15759} {\bibfield  {journal} {\bibinfo
  {journal} {Nature}\ }\textbf {\bibinfo {volume} {526}},\ \bibinfo {pages}
  {682} (\bibinfo {year} {2015})}\BibitemShut {NoStop}%
\bibitem [{\citenamefont {Togan}\ \emph {et~al.}(2010)\citenamefont {Togan},
  \citenamefont {Chu}, \citenamefont {Trifonov}, \citenamefont {Jiang},
  \citenamefont {Maze}, \citenamefont {Childress}, \citenamefont {Dutt},
  \citenamefont {S{\o}rensen}, \citenamefont {Hemmer}, \citenamefont {Zibrov},\
  and\ \citenamefont {Lukin}}]{Togan-SpinPhotonEntanglement-2010}%
  \BibitemOpen
  \bibfield  {author} {\bibinfo {author} {\bibfnamefont {E.}~\bibnamefont
  {Togan}}, \bibinfo {author} {\bibfnamefont {Y.}~\bibnamefont {Chu}}, \bibinfo
  {author} {\bibfnamefont {A.~S.}\ \bibnamefont {Trifonov}}, \bibinfo {author}
  {\bibfnamefont {L.}~\bibnamefont {Jiang}}, \bibinfo {author} {\bibfnamefont
  {J.}~\bibnamefont {Maze}}, \bibinfo {author} {\bibfnamefont {L.}~\bibnamefont
  {Childress}}, \bibinfo {author} {\bibfnamefont {M.~V.~G.}\ \bibnamefont
  {Dutt}}, \bibinfo {author} {\bibfnamefont {A.~S.}\ \bibnamefont
  {S{\o}rensen}}, \bibinfo {author} {\bibfnamefont {P.~R.}\ \bibnamefont
  {Hemmer}}, \bibinfo {author} {\bibfnamefont {A.~S.}\ \bibnamefont {Zibrov}},\
  and\ \bibinfo {author} {\bibfnamefont {M.~D.}\ \bibnamefont {Lukin}},\
  }\bibfield  {title} {\bibinfo {title} {{Quantum entanglement between an
  optical photon and a solid-state spin qubit.}},\ }\href
  {https://doi.org/10.1038/nature09256} {\bibfield  {journal} {\bibinfo
  {journal} {Nature}\ }\textbf {\bibinfo {volume} {466}},\ \bibinfo {pages}
  {730} (\bibinfo {year} {2010})}\BibitemShut {NoStop}%
\bibitem [{\citenamefont {Faraon}\ \emph {et~al.}(2012)\citenamefont {Faraon},
  \citenamefont {Santori}, \citenamefont {Huang}, \citenamefont {Acosta},\ and\
  \citenamefont {Beausoleil}}]{Faraon2012}%
  \BibitemOpen
  \bibfield  {author} {\bibinfo {author} {\bibfnamefont {A.}~\bibnamefont
  {Faraon}}, \bibinfo {author} {\bibfnamefont {C.}~\bibnamefont {Santori}},
  \bibinfo {author} {\bibfnamefont {Z.}~\bibnamefont {Huang}}, \bibinfo
  {author} {\bibfnamefont {V.~M.}\ \bibnamefont {Acosta}},\ and\ \bibinfo
  {author} {\bibfnamefont {R.~G.}\ \bibnamefont {Beausoleil}},\ }\bibfield
  {title} {\bibinfo {title} {{Coupling of Nitrogen-Vacancy Centers to Photonic
  Crystal Cavities in Monocrystalline Diamond}},\ }\href
  {https://doi.org/10.1103/PhysRevLett.109.033604} {\bibfield  {journal}
  {\bibinfo  {journal} {Physical Review Letters}\ }\textbf {\bibinfo {volume}
  {109}},\ \bibinfo {pages} {033604} (\bibinfo {year} {2012})}\BibitemShut
  {NoStop}%
\bibitem [{\citenamefont {Ruf}\ \emph {et~al.}(2020)\citenamefont {Ruf},
  \citenamefont {Weaver}, \citenamefont {van Dam},\ and\ \citenamefont
  {Hanson}}]{Ruf-FPCavity-arxiv-2020}%
  \BibitemOpen
  \bibfield  {author} {\bibinfo {author} {\bibfnamefont {M.}~\bibnamefont
  {Ruf}}, \bibinfo {author} {\bibfnamefont {M.~J.}\ \bibnamefont {Weaver}},
  \bibinfo {author} {\bibfnamefont {S.~B.}\ \bibnamefont {van Dam}},\ and\
  \bibinfo {author} {\bibfnamefont {R.}~\bibnamefont {Hanson}},\ }\href@noop {}
  {\bibinfo {title} {{Resonant Excitation and Purcell Enhancement of Coherent
  Nitrogen-Vacancy Centers Coupled to a Fabry-P{\'{e}}rot Micro-Cavity}}}
  (\bibinfo {year} {2020}),\ \Eprint {https://arxiv.org/abs/2009.08204}
  {arXiv:2009.08204 [quant-ph]} \BibitemShut {NoStop}%
\bibitem [{\citenamefont {Wang}\ and\ \citenamefont
  {Lekavicius}(2020)}]{wang-SpinMechanicsCoupling-2020-arxiv}%
  \BibitemOpen
  \bibfield  {author} {\bibinfo {author} {\bibfnamefont {H.}~\bibnamefont
  {Wang}}\ and\ \bibinfo {author} {\bibfnamefont {I.}~\bibnamefont
  {Lekavicius}},\ }\href@noop {} {\bibinfo {title} {{Coupling spins to
  nanomechanical resonators: Toward quantum spin-mechanics}}} (\bibinfo {year}
  {2020}),\ \Eprint {https://arxiv.org/abs/2011.09990} {arXiv:2011.09990
  [cond-mat.mes-hall]} \BibitemShut {NoStop}%
\bibitem [{\citenamefont {Lee}\ \emph {et~al.}(2017)\citenamefont {Lee},
  \citenamefont {Lee}, \citenamefont {Cady}, \citenamefont {Ovartchaiyapong},\
  and\ \citenamefont {Jayich}}]{Lee-SpinMechanicsReview-JoO-2017}%
  \BibitemOpen
  \bibfield  {author} {\bibinfo {author} {\bibfnamefont {D.}~\bibnamefont
  {Lee}}, \bibinfo {author} {\bibfnamefont {K.~W.}\ \bibnamefont {Lee}},
  \bibinfo {author} {\bibfnamefont {J.~V.}\ \bibnamefont {Cady}}, \bibinfo
  {author} {\bibfnamefont {P.}~\bibnamefont {Ovartchaiyapong}},\ and\ \bibinfo
  {author} {\bibfnamefont {A.~C.~B.}\ \bibnamefont {Jayich}},\ }\bibfield
  {title} {\bibinfo {title} {{Topical review: spins and mechanics in
  diamond}},\ }\href {https://doi.org/10.1088/2040-8986/aa52cd} {\bibfield
  {journal} {\bibinfo  {journal} {Journal of Optics}\ }\textbf {\bibinfo
  {volume} {19}},\ \bibinfo {pages} {33001} (\bibinfo {year}
  {2017})}\BibitemShut {NoStop}%
\bibitem [{\citenamefont {Aspelmeyer}\ \emph {et~al.}(2014)\citenamefont
  {Aspelmeyer}, \citenamefont {Kippenberg},\ and\ \citenamefont
  {Marquardt}}]{Kippenberg-2014-RMP-CavityOptomechanics}%
  \BibitemOpen
  \bibfield  {author} {\bibinfo {author} {\bibfnamefont {M.}~\bibnamefont
  {Aspelmeyer}}, \bibinfo {author} {\bibfnamefont {T.~J.}\ \bibnamefont
  {Kippenberg}},\ and\ \bibinfo {author} {\bibfnamefont {F.}~\bibnamefont
  {Marquardt}},\ }\bibfield  {title} {\bibinfo {title} {{Cavity
  optomechanics}},\ }\href {https://doi.org/10.1103/RevModPhys.86.1391}
  {\bibfield  {journal} {\bibinfo  {journal} {Reviews of Modern Physics}\
  }\textbf {\bibinfo {volume} {86}},\ \bibinfo {pages} {1391} (\bibinfo {year}
  {2014})}\BibitemShut {NoStop}%
\bibitem [{\citenamefont {Delsing}\ \emph {et~al.}(2019)\citenamefont
  {Delsing}, \citenamefont {Cleland}, \citenamefont {Schuetz}, \citenamefont
  {Kn{\"{o}}rzer}, \citenamefont {Giedke}, \citenamefont {Cirac}, \citenamefont
  {Srinivasan}, \citenamefont {Wu}, \citenamefont {Balram}, \citenamefont
  {B{\"{a}}uerle}, \citenamefont {Meunier}, \citenamefont {Ford}, \citenamefont
  {Santos}, \citenamefont {Cerda-M{\'{e}}ndez}, \citenamefont {Wang},
  \citenamefont {Krenner}, \citenamefont {Nysten}, \citenamefont {Nash},
  \citenamefont {Thevenard}, \citenamefont {Gourdon}, \citenamefont
  {Rovillain}, \citenamefont {Marangolo}, \citenamefont {Duquesne},
  \citenamefont {Fischerauer}, \citenamefont {Ruile}, \citenamefont {Reiner},
  \citenamefont {Paschke}, \citenamefont {Denysenko}, \citenamefont {Volkmer},
  \citenamefont {Wixforth}, \citenamefont {Bruus}, \citenamefont {Wiklund},
  \citenamefont {Reboud}, \citenamefont {Cooper}, \citenamefont {Fu},
  \citenamefont {Brugger}, \citenamefont {Rehfeldt},\ and\ \citenamefont
  {Westerhausen}}]{Delsing-SAWroadmap-JPhD-2019}%
  \BibitemOpen
  \bibfield  {author} {\bibinfo {author} {\bibfnamefont {P.}~\bibnamefont
  {Delsing}}, \bibinfo {author} {\bibfnamefont {A.~N.}\ \bibnamefont
  {Cleland}}, \bibinfo {author} {\bibfnamefont {M.~J.~A.}\ \bibnamefont
  {Schuetz}}, \bibinfo {author} {\bibfnamefont {J.}~\bibnamefont
  {Kn{\"{o}}rzer}}, \bibinfo {author} {\bibfnamefont {G.}~\bibnamefont
  {Giedke}}, \bibinfo {author} {\bibfnamefont {J.~I.}\ \bibnamefont {Cirac}},
  \bibinfo {author} {\bibfnamefont {K.}~\bibnamefont {Srinivasan}}, \bibinfo
  {author} {\bibfnamefont {M.}~\bibnamefont {Wu}}, \bibinfo {author}
  {\bibfnamefont {K.~C.}\ \bibnamefont {Balram}}, \bibinfo {author}
  {\bibfnamefont {C.}~\bibnamefont {B{\"{a}}uerle}}, \bibinfo {author}
  {\bibfnamefont {T.}~\bibnamefont {Meunier}}, \bibinfo {author} {\bibfnamefont
  {C.~J.~B.}\ \bibnamefont {Ford}}, \bibinfo {author} {\bibfnamefont {P.~V.}\
  \bibnamefont {Santos}}, \bibinfo {author} {\bibfnamefont {E.}~\bibnamefont
  {Cerda-M{\'{e}}ndez}}, \bibinfo {author} {\bibfnamefont {H.}~\bibnamefont
  {Wang}}, \bibinfo {author} {\bibfnamefont {H.~J.}\ \bibnamefont {Krenner}},
  \bibinfo {author} {\bibfnamefont {E.~D.~S.}\ \bibnamefont {Nysten}}, \bibinfo
  {author} {\bibfnamefont {M.~W. G.~R.}\ \bibnamefont {Nash}}, \bibinfo
  {author} {\bibfnamefont {L.}~\bibnamefont {Thevenard}}, \bibinfo {author}
  {\bibfnamefont {C.}~\bibnamefont {Gourdon}}, \bibinfo {author} {\bibfnamefont
  {P.}~\bibnamefont {Rovillain}}, \bibinfo {author} {\bibfnamefont
  {M.}~\bibnamefont {Marangolo}}, \bibinfo {author} {\bibfnamefont {J.-Y.}\
  \bibnamefont {Duquesne}}, \bibinfo {author} {\bibfnamefont {G.}~\bibnamefont
  {Fischerauer}}, \bibinfo {author} {\bibfnamefont {W.}~\bibnamefont {Ruile}},
  \bibinfo {author} {\bibfnamefont {A.}~\bibnamefont {Reiner}}, \bibinfo
  {author} {\bibfnamefont {B.}~\bibnamefont {Paschke}}, \bibinfo {author}
  {\bibfnamefont {D.}~\bibnamefont {Denysenko}}, \bibinfo {author}
  {\bibfnamefont {D.}~\bibnamefont {Volkmer}}, \bibinfo {author} {\bibfnamefont
  {A.}~\bibnamefont {Wixforth}}, \bibinfo {author} {\bibfnamefont
  {H.}~\bibnamefont {Bruus}}, \bibinfo {author} {\bibfnamefont
  {M.}~\bibnamefont {Wiklund}}, \bibinfo {author} {\bibfnamefont
  {J.}~\bibnamefont {Reboud}}, \bibinfo {author} {\bibfnamefont {J.~M.}\
  \bibnamefont {Cooper}}, \bibinfo {author} {\bibfnamefont {Y.}~\bibnamefont
  {Fu}}, \bibinfo {author} {\bibfnamefont {M.~S.}\ \bibnamefont {Brugger}},
  \bibinfo {author} {\bibfnamefont {F.}~\bibnamefont {Rehfeldt}},\ and\
  \bibinfo {author} {\bibfnamefont {C.}~\bibnamefont {Westerhausen}},\
  }\bibfield  {title} {\bibinfo {title} {{The 2019 surface acoustic waves
  roadmap}},\ }\href {https://doi.org/10.1088/1361-6463/ab1b04} {\bibfield
  {journal} {\bibinfo  {journal} {Journal of Physics D: Applied Physics}\
  }\textbf {\bibinfo {volume} {52}},\ \bibinfo {pages} {353001} (\bibinfo
  {year} {2019})}\BibitemShut {NoStop}%
\bibitem [{\citenamefont {Leibfried}\ \emph {et~al.}(2003)\citenamefont
  {Leibfried}, \citenamefont {Blatt}, \citenamefont {Monroe},\ and\
  \citenamefont {Wineland}}]{Leibfried-IonReview-RevModPhys-2003}%
  \BibitemOpen
  \bibfield  {author} {\bibinfo {author} {\bibfnamefont {D.}~\bibnamefont
  {Leibfried}}, \bibinfo {author} {\bibfnamefont {R.}~\bibnamefont {Blatt}},
  \bibinfo {author} {\bibfnamefont {C.}~\bibnamefont {Monroe}},\ and\ \bibinfo
  {author} {\bibfnamefont {D.}~\bibnamefont {Wineland}},\ }\bibfield  {title}
  {\bibinfo {title} {{Quantum dynamics of single trapped ions}},\ }\href
  {https://doi.org/10.1103/RevModPhys.75.281} {\bibfield  {journal} {\bibinfo
  {journal} {Rev. Mod. Phys.}\ }\textbf {\bibinfo {volume} {75}},\ \bibinfo
  {pages} {281} (\bibinfo {year} {2003})}\BibitemShut {NoStop}%
\bibitem [{\citenamefont {Bienfait}\ \emph {et~al.}(2019)\citenamefont
  {Bienfait}, \citenamefont {Satzinger}, \citenamefont {Zhong}, \citenamefont
  {Chang}, \citenamefont {Chou}, \citenamefont {Conner}, \citenamefont {Dumur},
  \citenamefont {Grebel}, \citenamefont {Peairs}, \citenamefont {Povey},\ and\
  \citenamefont {Cleland}}]{Bienfait-PhononSCQubit-Science-2019}%
  \BibitemOpen
  \bibfield  {author} {\bibinfo {author} {\bibfnamefont {A.}~\bibnamefont
  {Bienfait}}, \bibinfo {author} {\bibfnamefont {K.~J.}\ \bibnamefont
  {Satzinger}}, \bibinfo {author} {\bibfnamefont {Y.~P.}\ \bibnamefont
  {Zhong}}, \bibinfo {author} {\bibfnamefont {H.-S.}\ \bibnamefont {Chang}},
  \bibinfo {author} {\bibfnamefont {M.-H.}\ \bibnamefont {Chou}}, \bibinfo
  {author} {\bibfnamefont {C.~R.}\ \bibnamefont {Conner}}, \bibinfo {author}
  {\bibfnamefont {{\'{E}}.}~\bibnamefont {Dumur}}, \bibinfo {author}
  {\bibfnamefont {J.}~\bibnamefont {Grebel}}, \bibinfo {author} {\bibfnamefont
  {G.~A.}\ \bibnamefont {Peairs}}, \bibinfo {author} {\bibfnamefont {R.~G.}\
  \bibnamefont {Povey}},\ and\ \bibinfo {author} {\bibfnamefont {A.~N.}\
  \bibnamefont {Cleland}},\ }\bibfield  {title} {\bibinfo {title}
  {{Phonon-mediated quantum state transfer and remote qubit entanglement}},\
  }\href {https://doi.org/10.1126/science.aaw8415} {\bibfield  {journal}
  {\bibinfo  {journal} {Science}\ }\textbf {\bibinfo {volume} {364}},\ \bibinfo
  {pages} {368} (\bibinfo {year} {2019})}\BibitemShut {NoStop}%
\bibitem [{\citenamefont {Macquarrie}\ \emph {et~al.}(2013)\citenamefont
  {Macquarrie}, \citenamefont {Gosavi}, \citenamefont {Jungwirth},
  \citenamefont {Bhave},\ and\ \citenamefont
  {Fuchs}}]{MacQuarrie-PRL-2013-StrainNVControl}%
  \BibitemOpen
  \bibfield  {author} {\bibinfo {author} {\bibfnamefont {E.~R.}\ \bibnamefont
  {Macquarrie}}, \bibinfo {author} {\bibfnamefont {T.~A.}\ \bibnamefont
  {Gosavi}}, \bibinfo {author} {\bibfnamefont {N.~R.}\ \bibnamefont
  {Jungwirth}}, \bibinfo {author} {\bibfnamefont {S.~A.}\ \bibnamefont
  {Bhave}},\ and\ \bibinfo {author} {\bibfnamefont {G.~D.}\ \bibnamefont
  {Fuchs}},\ }\bibfield  {title} {\bibinfo {title} {{Mechanical spin control of
  nitrogen-vacancy centers in diamond}},\ }\href
  {https://doi.org/10.1103/PhysRevLett.111.227602} {\bibfield  {journal}
  {\bibinfo  {journal} {Physical Review Letters}\ }\textbf {\bibinfo {volume}
  {111}},\ \bibinfo {pages} {227602} (\bibinfo {year} {2013})},\ \Eprint
  {https://arxiv.org/abs/1306.6356} {arXiv:1306.6356} \BibitemShut {NoStop}%
\bibitem [{\citenamefont {Golter}\ \emph {et~al.}(2016)\citenamefont {Golter},
  \citenamefont {Oo}, \citenamefont {Amezcua}, \citenamefont {Lekavicius},
  \citenamefont {Stewart},\ and\ \citenamefont
  {Wang}}]{Golter-PRX-2016-NVcoupledSAW}%
  \BibitemOpen
  \bibfield  {author} {\bibinfo {author} {\bibfnamefont {D.~A.}\ \bibnamefont
  {Golter}}, \bibinfo {author} {\bibfnamefont {T.}~\bibnamefont {Oo}}, \bibinfo
  {author} {\bibfnamefont {M.}~\bibnamefont {Amezcua}}, \bibinfo {author}
  {\bibfnamefont {I.}~\bibnamefont {Lekavicius}}, \bibinfo {author}
  {\bibfnamefont {K.~A.}\ \bibnamefont {Stewart}},\ and\ \bibinfo {author}
  {\bibfnamefont {H.}~\bibnamefont {Wang}},\ }\bibfield  {title} {\bibinfo
  {title} {{Coupling a Surface Acoustic Wave to an Electron Spin in Diamond via
  a Dark State}},\ }\href {https://doi.org/10.1103/PhysRevX.6.041060}
  {\bibfield  {journal} {\bibinfo  {journal} {Physical Review X}\ }\textbf
  {\bibinfo {volume} {6}},\ \bibinfo {pages} {41060} (\bibinfo {year}
  {2016})}\BibitemShut {NoStop}%
\bibitem [{\citenamefont {Whiteley}\ \emph {et~al.}(2019)\citenamefont
  {Whiteley}, \citenamefont {Wolfowicz}, \citenamefont {Anderson},
  \citenamefont {Bourassa}, \citenamefont {Ma}, \citenamefont {Ye},
  \citenamefont {Koolstra}, \citenamefont {Satzinger}, \citenamefont {Holt},
  \citenamefont {Heremans}, \citenamefont {Cleland}, \citenamefont {Schuster},
  \citenamefont {Galli},\ and\ \citenamefont
  {Awschalom}}]{Whiteley-SiCspin-phononCoupling-NatPhys-2019}%
  \BibitemOpen
  \bibfield  {author} {\bibinfo {author} {\bibfnamefont {S.~J.}\ \bibnamefont
  {Whiteley}}, \bibinfo {author} {\bibfnamefont {G.}~\bibnamefont {Wolfowicz}},
  \bibinfo {author} {\bibfnamefont {C.~P.}\ \bibnamefont {Anderson}}, \bibinfo
  {author} {\bibfnamefont {A.}~\bibnamefont {Bourassa}}, \bibinfo {author}
  {\bibfnamefont {H.}~\bibnamefont {Ma}}, \bibinfo {author} {\bibfnamefont
  {M.}~\bibnamefont {Ye}}, \bibinfo {author} {\bibfnamefont {G.}~\bibnamefont
  {Koolstra}}, \bibinfo {author} {\bibfnamefont {K.~J.}\ \bibnamefont
  {Satzinger}}, \bibinfo {author} {\bibfnamefont {M.~V.}\ \bibnamefont {Holt}},
  \bibinfo {author} {\bibfnamefont {F.~J.}\ \bibnamefont {Heremans}}, \bibinfo
  {author} {\bibfnamefont {A.~N.}\ \bibnamefont {Cleland}}, \bibinfo {author}
  {\bibfnamefont {D.~I.}\ \bibnamefont {Schuster}}, \bibinfo {author}
  {\bibfnamefont {G.}~\bibnamefont {Galli}},\ and\ \bibinfo {author}
  {\bibfnamefont {D.~D.}\ \bibnamefont {Awschalom}},\ }\bibfield  {title}
  {\bibinfo {title} {{Spin–phonon interactions in silicon carbide addressed
  by Gaussian acoustics}},\ }\href {https://doi.org/10.1038/s41567-019-0420-0}
  {\bibfield  {journal} {\bibinfo  {journal} {Nature Physics}\ }\textbf
  {\bibinfo {volume} {15}},\ \bibinfo {pages} {490} (\bibinfo {year}
  {2019})}\BibitemShut {NoStop}%
\bibitem [{\citenamefont {Maity}\ \emph {et~al.}(2020)\citenamefont {Maity},
  \citenamefont {Shao}, \citenamefont {Bogdanovi{\'{c}}}, \citenamefont
  {Meesala}, \citenamefont {Sohn}, \citenamefont {Sinclair}, \citenamefont
  {Pingault}, \citenamefont {Chalupnik}, \citenamefont {Chia}, \citenamefont
  {Zheng}, \citenamefont {Lai},\ and\ \citenamefont
  {Lon{\v{c}}ar}}]{Maity-2020-NatComm-CoherentContolSiVacoustic}%
  \BibitemOpen
  \bibfield  {author} {\bibinfo {author} {\bibfnamefont {S.}~\bibnamefont
  {Maity}}, \bibinfo {author} {\bibfnamefont {L.}~\bibnamefont {Shao}},
  \bibinfo {author} {\bibfnamefont {S.}~\bibnamefont {Bogdanovi{\'{c}}}},
  \bibinfo {author} {\bibfnamefont {S.}~\bibnamefont {Meesala}}, \bibinfo
  {author} {\bibfnamefont {Y.-I.}\ \bibnamefont {Sohn}}, \bibinfo {author}
  {\bibfnamefont {N.}~\bibnamefont {Sinclair}}, \bibinfo {author}
  {\bibfnamefont {B.}~\bibnamefont {Pingault}}, \bibinfo {author}
  {\bibfnamefont {M.}~\bibnamefont {Chalupnik}}, \bibinfo {author}
  {\bibfnamefont {C.}~\bibnamefont {Chia}}, \bibinfo {author} {\bibfnamefont
  {L.}~\bibnamefont {Zheng}}, \bibinfo {author} {\bibfnamefont
  {K.}~\bibnamefont {Lai}},\ and\ \bibinfo {author} {\bibfnamefont
  {M.}~\bibnamefont {Lon{\v{c}}ar}},\ }\bibfield  {title} {\bibinfo {title}
  {{Coherent acoustic control of a single silicon vacancy spin in diamond}},\
  }\href {https://doi.org/10.1038/s41467-019-13822-x} {\bibfield  {journal}
  {\bibinfo  {journal} {Nature Communications}\ }\textbf {\bibinfo {volume}
  {11}},\ \bibinfo {pages} {193} (\bibinfo {year} {2020})}\BibitemShut
  {NoStop}%
\bibitem [{\citenamefont {Meesala}\ \emph {et~al.}(2016)\citenamefont
  {Meesala}, \citenamefont {Sohn}, \citenamefont {Atikian}, \citenamefont
  {Kim}, \citenamefont {Burek}, \citenamefont {Choy},\ and\ \citenamefont
  {Lon{\v{c}}ar}}]{Mesala-2016-PRAppl-NVmechanics}%
  \BibitemOpen
  \bibfield  {author} {\bibinfo {author} {\bibfnamefont {S.}~\bibnamefont
  {Meesala}}, \bibinfo {author} {\bibfnamefont {Y.-I.}\ \bibnamefont {Sohn}},
  \bibinfo {author} {\bibfnamefont {H.~A.}\ \bibnamefont {Atikian}}, \bibinfo
  {author} {\bibfnamefont {S.}~\bibnamefont {Kim}}, \bibinfo {author}
  {\bibfnamefont {M.~J.}\ \bibnamefont {Burek}}, \bibinfo {author}
  {\bibfnamefont {J.~T.}\ \bibnamefont {Choy}},\ and\ \bibinfo {author}
  {\bibfnamefont {M.}~\bibnamefont {Lon{\v{c}}ar}},\ }\bibfield  {title}
  {\bibinfo {title} {{Enhanced Strain Coupling of Nitrogen-Vacancy Spins to
  Nanoscale Diamond Cantilevers}},\ }\href
  {https://doi.org/10.1103/PhysRevApplied.5.034010} {\bibfield  {journal}
  {\bibinfo  {journal} {Physical Review Applied}\ }\textbf {\bibinfo {volume}
  {5}},\ \bibinfo {pages} {34010} (\bibinfo {year} {2016})}\BibitemShut
  {NoStop}%
\bibitem [{\citenamefont {Ovartchaiyapong}\ \emph {et~al.}(2014)\citenamefont
  {Ovartchaiyapong}, \citenamefont {Lee}, \citenamefont {Myers},\ and\
  \citenamefont {Jayich}}]{Ovartchaiyapong2014}%
  \BibitemOpen
  \bibfield  {author} {\bibinfo {author} {\bibfnamefont {P.}~\bibnamefont
  {Ovartchaiyapong}}, \bibinfo {author} {\bibfnamefont {K.~W.}\ \bibnamefont
  {Lee}}, \bibinfo {author} {\bibfnamefont {B.~A.}\ \bibnamefont {Myers}},\
  and\ \bibinfo {author} {\bibfnamefont {A.~C.~B.}\ \bibnamefont {Jayich}},\
  }\bibfield  {title} {\bibinfo {title} {{Dynamic strain-mediated coupling of a
  single diamond spin to a mechanical resonator.}},\ }\href
  {https://doi.org/10.1038/ncomms5429} {\bibfield  {journal} {\bibinfo
  {journal} {Nature communications}\ }\textbf {\bibinfo {volume} {5}},\
  \bibinfo {pages} {4429} (\bibinfo {year} {2014})}\BibitemShut {NoStop}%
\bibitem [{\citenamefont {Barfuss}\ \emph {et~al.}(2015)\citenamefont
  {Barfuss}, \citenamefont {Teissier}, \citenamefont {Neu}, \citenamefont
  {Nunnenkamp},\ and\ \citenamefont
  {Maletinsky}}]{Barfuss-StrongMechanicalDrivinfNV-NatPhys-2015}%
  \BibitemOpen
  \bibfield  {author} {\bibinfo {author} {\bibfnamefont {A.}~\bibnamefont
  {Barfuss}}, \bibinfo {author} {\bibfnamefont {J.}~\bibnamefont {Teissier}},
  \bibinfo {author} {\bibfnamefont {E.}~\bibnamefont {Neu}}, \bibinfo {author}
  {\bibfnamefont {A.}~\bibnamefont {Nunnenkamp}},\ and\ \bibinfo {author}
  {\bibfnamefont {P.}~\bibnamefont {Maletinsky}},\ }\bibfield  {title}
  {\bibinfo {title} {{Strong mechanical driving of a single electron spin}},\
  }\href {https://doi.org/10.1038/nphys3411} {\bibfield  {journal} {\bibinfo
  {journal} {Nature Physics}\ }\textbf {\bibinfo {volume} {11}},\ \bibinfo
  {pages} {820} (\bibinfo {year} {2015})}\BibitemShut {NoStop}%
\bibitem [{\citenamefont {Arcizet}\ \emph {et~al.}(2011)\citenamefont
  {Arcizet}, \citenamefont {Jacques}, \citenamefont {Siria}, \citenamefont
  {Poncharal}, \citenamefont {Vincent},\ and\ \citenamefont
  {Seidelin}}]{Arcizet-NVNanowire-2011-NatPhys}%
  \BibitemOpen
  \bibfield  {author} {\bibinfo {author} {\bibfnamefont {O.}~\bibnamefont
  {Arcizet}}, \bibinfo {author} {\bibfnamefont {V.}~\bibnamefont {Jacques}},
  \bibinfo {author} {\bibfnamefont {A.}~\bibnamefont {Siria}}, \bibinfo
  {author} {\bibfnamefont {P.}~\bibnamefont {Poncharal}}, \bibinfo {author}
  {\bibfnamefont {P.}~\bibnamefont {Vincent}},\ and\ \bibinfo {author}
  {\bibfnamefont {S.}~\bibnamefont {Seidelin}},\ }\bibfield  {title} {\bibinfo
  {title} {{A single nitrogen-vacancy defect coupled to a nanomechanical
  oscillator}},\ }\href {https://doi.org/10.1038/nphys2070} {\bibfield
  {journal} {\bibinfo  {journal} {Nature Physics}\ }\textbf {\bibinfo {volume}
  {7}},\ \bibinfo {pages} {879} (\bibinfo {year} {2011})}\BibitemShut {NoStop}%
\bibitem [{\citenamefont {Pigeau}\ \emph {et~al.}(2015)\citenamefont {Pigeau},
  \citenamefont {Rohr}, \citenamefont {{Mercier De Lepinay}}, \citenamefont
  {Gloppe}, \citenamefont {Jacques},\ and\ \citenamefont
  {Arcizet}}]{Pigeau-NVcouledToNanowireSpinProtection-NatCom-2015}%
  \BibitemOpen
  \bibfield  {author} {\bibinfo {author} {\bibfnamefont {B.}~\bibnamefont
  {Pigeau}}, \bibinfo {author} {\bibfnamefont {S.}~\bibnamefont {Rohr}},
  \bibinfo {author} {\bibfnamefont {L.}~\bibnamefont {{Mercier De Lepinay}}},
  \bibinfo {author} {\bibfnamefont {A.}~\bibnamefont {Gloppe}}, \bibinfo
  {author} {\bibfnamefont {V.}~\bibnamefont {Jacques}},\ and\ \bibinfo {author}
  {\bibfnamefont {O.}~\bibnamefont {Arcizet}},\ }\bibfield  {title} {\bibinfo
  {title} {{Observation of a phononic Mollow triplet in a multimode hybrid
  spin-nanomechanical system}},\ }\href {https://doi.org/10.1038/ncomms9603}
  {\bibfield  {journal} {\bibinfo  {journal} {Nature Communications}\ }\textbf
  {\bibinfo {volume} {6}},\ \bibinfo {pages} {1} (\bibinfo {year}
  {2015})}\BibitemShut {NoStop}%
\bibitem [{\citenamefont {Ohta}\ \emph {et~al.}(2021)\citenamefont {Ohta},
  \citenamefont {Herpin}, \citenamefont {Bastidas}, \citenamefont {Tawara},
  \citenamefont {Yamaguchi},\ and\ \citenamefont
  {Okamoto}}]{Ohta-OptomechanicsREI-PRL-2021}%
  \BibitemOpen
  \bibfield  {author} {\bibinfo {author} {\bibfnamefont {R.}~\bibnamefont
  {Ohta}}, \bibinfo {author} {\bibfnamefont {L.}~\bibnamefont {Herpin}},
  \bibinfo {author} {\bibfnamefont {V.~M.}\ \bibnamefont {Bastidas}}, \bibinfo
  {author} {\bibfnamefont {T.}~\bibnamefont {Tawara}}, \bibinfo {author}
  {\bibfnamefont {H.}~\bibnamefont {Yamaguchi}},\ and\ \bibinfo {author}
  {\bibfnamefont {H.}~\bibnamefont {Okamoto}},\ }\bibfield  {title} {\bibinfo
  {title} {{Rare-Earth-Mediated Optomechanical System in the Reversed
  Dissipation Regime}},\ }\href
  {https://doi.org/10.1103/PhysRevLett.126.047404} {\bibfield  {journal}
  {\bibinfo  {journal} {Phys. Rev. Lett.}\ }\textbf {\bibinfo {volume} {126}},\
  \bibinfo {pages} {47404} (\bibinfo {year} {2021})}\BibitemShut {NoStop}%
\bibitem [{\citenamefont {Burek}\ \emph {et~al.}(2016)\citenamefont {Burek},
  \citenamefont {Cohen}, \citenamefont {Meenehan}, \citenamefont {El-Sawah},
  \citenamefont {Chia}, \citenamefont {Ruelle}, \citenamefont {Meesala},
  \citenamefont {Rochman}, \citenamefont {Atikian}, \citenamefont {Markham},
  \citenamefont {Twitchen}, \citenamefont {Lukin}, \citenamefont {Painter},\
  and\ \citenamefont
  {Lon{\v{c}}ar}}]{Burek-DiamondOptomechanicalCrystal-Optica-2016}%
  \BibitemOpen
  \bibfield  {author} {\bibinfo {author} {\bibfnamefont {M.~J.}\ \bibnamefont
  {Burek}}, \bibinfo {author} {\bibfnamefont {J.~D.}\ \bibnamefont {Cohen}},
  \bibinfo {author} {\bibfnamefont {S.~M.}\ \bibnamefont {Meenehan}}, \bibinfo
  {author} {\bibfnamefont {N.}~\bibnamefont {El-Sawah}}, \bibinfo {author}
  {\bibfnamefont {C.}~\bibnamefont {Chia}}, \bibinfo {author} {\bibfnamefont
  {T.}~\bibnamefont {Ruelle}}, \bibinfo {author} {\bibfnamefont
  {S.}~\bibnamefont {Meesala}}, \bibinfo {author} {\bibfnamefont
  {J.}~\bibnamefont {Rochman}}, \bibinfo {author} {\bibfnamefont {H.~A.}\
  \bibnamefont {Atikian}}, \bibinfo {author} {\bibfnamefont {M.}~\bibnamefont
  {Markham}}, \bibinfo {author} {\bibfnamefont {D.~J.}\ \bibnamefont
  {Twitchen}}, \bibinfo {author} {\bibfnamefont {M.~D.}\ \bibnamefont {Lukin}},
  \bibinfo {author} {\bibfnamefont {O.}~\bibnamefont {Painter}},\ and\ \bibinfo
  {author} {\bibfnamefont {M.}~\bibnamefont {Lon{\v{c}}ar}},\ }\bibfield
  {title} {\bibinfo {title} {{Diamond optomechanical crystals}},\ }\href
  {https://doi.org/10.1364/OPTICA.3.001404} {\bibfield  {journal} {\bibinfo
  {journal} {Optica}\ }\textbf {\bibinfo {volume} {3}},\ \bibinfo {pages}
  {1404} (\bibinfo {year} {2016})}\BibitemShut {NoStop}%
\bibitem [{\citenamefont {Mitchell}\ \emph {et~al.}(2016)\citenamefont
  {Mitchell}, \citenamefont {Khanaliloo}, \citenamefont {Lake}, \citenamefont
  {Masuda}, \citenamefont {Hadden},\ and\ \citenamefont
  {Barclay}}]{Mitchell-DiamondOptomechanicalResonator-Optica-2016}%
  \BibitemOpen
  \bibfield  {author} {\bibinfo {author} {\bibfnamefont {M.}~\bibnamefont
  {Mitchell}}, \bibinfo {author} {\bibfnamefont {B.}~\bibnamefont
  {Khanaliloo}}, \bibinfo {author} {\bibfnamefont {D.~P.}\ \bibnamefont
  {Lake}}, \bibinfo {author} {\bibfnamefont {T.}~\bibnamefont {Masuda}},
  \bibinfo {author} {\bibfnamefont {J.~P.}\ \bibnamefont {Hadden}},\ and\
  \bibinfo {author} {\bibfnamefont {P.~E.}\ \bibnamefont {Barclay}},\
  }\bibfield  {title} {\bibinfo {title} {{Single-crystal diamond
  low-dissipation cavity optomechanics}},\ }\href
  {https://doi.org/10.1364/OPTICA.3.000963} {\bibfield  {journal} {\bibinfo
  {journal} {Optica}\ }\textbf {\bibinfo {volume} {3}},\ \bibinfo {pages} {963}
  (\bibinfo {year} {2016})}\BibitemShut {NoStop}%
\bibitem [{\citenamefont {Cohen}\ \emph {et~al.}(2015)\citenamefont {Cohen},
  \citenamefont {Meenehan}, \citenamefont {MacCabe}, \citenamefont
  {Gr{\"{o}}blacher}, \citenamefont {Safavi-Naeini}, \citenamefont {Marsili},
  \citenamefont {Shaw},\ and\ \citenamefont
  {Painter}}]{Cohen-PhononCounting-Nature-2015}%
  \BibitemOpen
  \bibfield  {author} {\bibinfo {author} {\bibfnamefont {J.~D.}\ \bibnamefont
  {Cohen}}, \bibinfo {author} {\bibfnamefont {S.~M.}\ \bibnamefont {Meenehan}},
  \bibinfo {author} {\bibfnamefont {G.~S.}\ \bibnamefont {MacCabe}}, \bibinfo
  {author} {\bibfnamefont {S.}~\bibnamefont {Gr{\"{o}}blacher}}, \bibinfo
  {author} {\bibfnamefont {A.~H.}\ \bibnamefont {Safavi-Naeini}}, \bibinfo
  {author} {\bibfnamefont {F.}~\bibnamefont {Marsili}}, \bibinfo {author}
  {\bibfnamefont {M.~D.}\ \bibnamefont {Shaw}},\ and\ \bibinfo {author}
  {\bibfnamefont {O.}~\bibnamefont {Painter}},\ }\bibfield  {title} {\bibinfo
  {title} {{Phonon counting and intensity interferometry of a nanomechanical
  resonator}},\ }\href {https://doi.org/10.1038/nature14349} {\bibfield
  {journal} {\bibinfo  {journal} {Nature}\ }\textbf {\bibinfo {volume} {520}},\
  \bibinfo {pages} {522} (\bibinfo {year} {2015})}\BibitemShut {NoStop}%
\bibitem [{\citenamefont {Wallucks}\ \emph {et~al.}(2020)\citenamefont
  {Wallucks}, \citenamefont {Marinkovi{\'{c}}}, \citenamefont {Hensen},
  \citenamefont {Stockill},\ and\ \citenamefont
  {Gr{\"{o}}blacher}}]{Wallucks-QuantumMemoryTelecomBand-NatPhys-2020}%
  \BibitemOpen
  \bibfield  {author} {\bibinfo {author} {\bibfnamefont {A.}~\bibnamefont
  {Wallucks}}, \bibinfo {author} {\bibfnamefont {I.}~\bibnamefont
  {Marinkovi{\'{c}}}}, \bibinfo {author} {\bibfnamefont {B.}~\bibnamefont
  {Hensen}}, \bibinfo {author} {\bibfnamefont {R.}~\bibnamefont {Stockill}},\
  and\ \bibinfo {author} {\bibfnamefont {S.}~\bibnamefont {Gr{\"{o}}blacher}},\
  }\bibfield  {title} {\bibinfo {title} {{A quantum memory at telecom
  wavelengths}},\ }\href {https://doi.org/10.1038/s41567-020-0891-z} {\bibfield
   {journal} {\bibinfo  {journal} {Nature Physics}\ }\textbf {\bibinfo {volume}
  {16}},\ \bibinfo {pages} {772} (\bibinfo {year} {2020})}\BibitemShut
  {NoStop}%
\bibitem [{\citenamefont {Regal}\ and\ \citenamefont
  {Lehnert}(2011)}]{Regal-QuantumTransducer-JoP-2011}%
  \BibitemOpen
  \bibfield  {author} {\bibinfo {author} {\bibfnamefont {C.~A.}\ \bibnamefont
  {Regal}}\ and\ \bibinfo {author} {\bibfnamefont {K.~W.}\ \bibnamefont
  {Lehnert}},\ }\bibfield  {title} {\bibinfo {title} {{From cavity
  electromechanics to cavity optomechanics}},\ }\href
  {https://doi.org/10.1088/1742-6596/264/1/012025} {\bibfield  {journal}
  {\bibinfo  {journal} {Journal of Physics: Conference Series}\ }\textbf
  {\bibinfo {volume} {264}},\ \bibinfo {pages} {12025} (\bibinfo {year}
  {2011})}\BibitemShut {NoStop}%
\bibitem [{\citenamefont {Mirhosseini}\ \emph {et~al.}(2020)\citenamefont
  {Mirhosseini}, \citenamefont {Sipahigil}, \citenamefont {Kalaee},\ and\
  \citenamefont {Painter}}]{Mirhosseini-arxiv-2020-QuantumTransduction}%
  \BibitemOpen
  \bibfield  {author} {\bibinfo {author} {\bibfnamefont {M.}~\bibnamefont
  {Mirhosseini}}, \bibinfo {author} {\bibfnamefont {A.}~\bibnamefont
  {Sipahigil}}, \bibinfo {author} {\bibfnamefont {M.}~\bibnamefont {Kalaee}},\
  and\ \bibinfo {author} {\bibfnamefont {O.}~\bibnamefont {Painter}},\
  }\bibfield  {title} {\bibinfo {title} {{Quantum transduction of optical
  photons from a superconducting qubit}},\ }\href
  {http://arxiv.org/abs/2004.04838} {\bibfield  {journal} {\bibinfo  {journal}
  {arXiv:}\ ,\ \bibinfo {pages} {2004.04838}} (\bibinfo {year} {2020})},\
  \Eprint {https://arxiv.org/abs/2004.04838} {arXiv:2004.04838} \BibitemShut
  {NoStop}%
\bibitem [{\citenamefont {Forsch}\ \emph {et~al.}(2020)\citenamefont {Forsch},
  \citenamefont {Stockill}, \citenamefont {Wallucks}, \citenamefont
  {Marinkovi{\'{c}}}, \citenamefont {G{\"{a}}rtner}, \citenamefont {Norte},
  \citenamefont {van Otten}, \citenamefont {Fiore}, \citenamefont
  {Srinivasan},\ and\ \citenamefont
  {Gr{\"{o}}blacher}}]{Forsh-MwtoOpticsTransduvtionOptomechanics-NatPhys-2020}%
  \BibitemOpen
  \bibfield  {author} {\bibinfo {author} {\bibfnamefont {M.}~\bibnamefont
  {Forsch}}, \bibinfo {author} {\bibfnamefont {R.}~\bibnamefont {Stockill}},
  \bibinfo {author} {\bibfnamefont {A.}~\bibnamefont {Wallucks}}, \bibinfo
  {author} {\bibfnamefont {I.}~\bibnamefont {Marinkovi{\'{c}}}}, \bibinfo
  {author} {\bibfnamefont {C.}~\bibnamefont {G{\"{a}}rtner}}, \bibinfo {author}
  {\bibfnamefont {R.~A.}\ \bibnamefont {Norte}}, \bibinfo {author}
  {\bibfnamefont {F.}~\bibnamefont {van Otten}}, \bibinfo {author}
  {\bibfnamefont {A.}~\bibnamefont {Fiore}}, \bibinfo {author} {\bibfnamefont
  {K.}~\bibnamefont {Srinivasan}},\ and\ \bibinfo {author} {\bibfnamefont
  {S.}~\bibnamefont {Gr{\"{o}}blacher}},\ }\bibfield  {title} {\bibinfo {title}
  {{Microwave-to-optics conversion using a mechanical oscillator in its quantum
  ground state}},\ }\href {https://doi.org/10.1038/s41567-019-0673-7}
  {\bibfield  {journal} {\bibinfo  {journal} {Nature Physics}\ }\textbf
  {\bibinfo {volume} {16}},\ \bibinfo {pages} {69} (\bibinfo {year}
  {2020})}\BibitemShut {NoStop}%
\bibitem [{\citenamefont {Lauk}\ \emph {et~al.}(2020)\citenamefont {Lauk},
  \citenamefont {Sinclair}, \citenamefont {Barzanjeh}, \citenamefont {Covey},
  \citenamefont {Saffman}, \citenamefont {Spiropulu},\ and\ \citenamefont
  {Simon}}]{Lauk-QuantumTransduction-QScTech-2020}%
  \BibitemOpen
  \bibfield  {author} {\bibinfo {author} {\bibfnamefont {N.}~\bibnamefont
  {Lauk}}, \bibinfo {author} {\bibfnamefont {N.}~\bibnamefont {Sinclair}},
  \bibinfo {author} {\bibfnamefont {S.}~\bibnamefont {Barzanjeh}}, \bibinfo
  {author} {\bibfnamefont {J.~P.}\ \bibnamefont {Covey}}, \bibinfo {author}
  {\bibfnamefont {M.}~\bibnamefont {Saffman}}, \bibinfo {author} {\bibfnamefont
  {M.}~\bibnamefont {Spiropulu}},\ and\ \bibinfo {author} {\bibfnamefont
  {C.}~\bibnamefont {Simon}},\ }\bibfield  {title} {\bibinfo {title}
  {{Perspectives on quantum transduction}},\ }\href
  {https://doi.org/10.1088/2058-9565/ab788a} {\bibfield  {journal} {\bibinfo
  {journal} {Quantum Science and Technology}\ }\textbf {\bibinfo {volume}
  {5}},\ \bibinfo {pages} {020501} (\bibinfo {year} {2020})}\BibitemShut
  {NoStop}%
\bibitem [{\citenamefont {Doherty}\ \emph {et~al.}(2013)\citenamefont
  {Doherty}, \citenamefont {Manson}, \citenamefont {Delaney}, \citenamefont
  {Jelezko}, \citenamefont {Wrachtrup},\ and\ \citenamefont
  {Hollenberg}}]{Doherty2013}%
  \BibitemOpen
  \bibfield  {author} {\bibinfo {author} {\bibfnamefont {M.~W.}\ \bibnamefont
  {Doherty}}, \bibinfo {author} {\bibfnamefont {N.~B.}\ \bibnamefont {Manson}},
  \bibinfo {author} {\bibfnamefont {P.}~\bibnamefont {Delaney}}, \bibinfo
  {author} {\bibfnamefont {F.}~\bibnamefont {Jelezko}}, \bibinfo {author}
  {\bibfnamefont {J.}~\bibnamefont {Wrachtrup}},\ and\ \bibinfo {author}
  {\bibfnamefont {L.~C.}\ \bibnamefont {Hollenberg}},\ }\href
  {https://doi.org/10.1016/j.physrep.2013.02.001} {\bibinfo {title} {{The
  nitrogen-vacancy colour centre in diamond}}} (\bibinfo {year} {2013}),\
  \Eprint {https://arxiv.org/abs/1302.3288} {arXiv:1302.3288} \BibitemShut
  {NoStop}%
\bibitem [{\citenamefont {Soykal}\ \emph {et~al.}(2011)\citenamefont {Soykal},
  \citenamefont {Ruskov},\ and\ \citenamefont
  {Tahan}}]{Soykal-QubitsSiliconPhonons-PRL-2011}%
  \BibitemOpen
  \bibfield  {author} {\bibinfo {author} {\bibfnamefont {{\"{O}}.~O.}\
  \bibnamefont {Soykal}}, \bibinfo {author} {\bibfnamefont {R.}~\bibnamefont
  {Ruskov}},\ and\ \bibinfo {author} {\bibfnamefont {C.}~\bibnamefont
  {Tahan}},\ }\bibfield  {title} {\bibinfo {title} {{Sound-Based Analogue of
  Cavity Quantum Electrodynamics in Silicon}},\ }\href
  {https://doi.org/10.1103/PhysRevLett.107.235502} {\bibfield  {journal}
  {\bibinfo  {journal} {Phys. Rev. Lett.}\ }\textbf {\bibinfo {volume} {107}},\
  \bibinfo {pages} {235502} (\bibinfo {year} {2011})}\BibitemShut {NoStop}%
\bibitem [{\citenamefont {Yeo}\ \emph {et~al.}(2014)\citenamefont {Yeo},
  \citenamefont {de~Assis}, \citenamefont {Gloppe}, \citenamefont
  {Dupont-Ferrier}, \citenamefont {Verlot}, \citenamefont {Malik},
  \citenamefont {Dupuy}, \citenamefont {Claudon}, \citenamefont {G{\'{e}}rard},
  \citenamefont {Auff{\`{e}}ves}, \citenamefont {Nogues}, \citenamefont
  {Seidelin}, \citenamefont {Poizat}, \citenamefont {Arcizet},\ and\
  \citenamefont {Richard}}]{Yeo--QD-phonon-coupling--NNano-2014}%
  \BibitemOpen
  \bibfield  {author} {\bibinfo {author} {\bibfnamefont {I.}~\bibnamefont
  {Yeo}}, \bibinfo {author} {\bibfnamefont {P.-L.}\ \bibnamefont {de~Assis}},
  \bibinfo {author} {\bibfnamefont {A.}~\bibnamefont {Gloppe}}, \bibinfo
  {author} {\bibfnamefont {E.}~\bibnamefont {Dupont-Ferrier}}, \bibinfo
  {author} {\bibfnamefont {P.}~\bibnamefont {Verlot}}, \bibinfo {author}
  {\bibfnamefont {N.~S.}\ \bibnamefont {Malik}}, \bibinfo {author}
  {\bibfnamefont {E.}~\bibnamefont {Dupuy}}, \bibinfo {author} {\bibfnamefont
  {J.}~\bibnamefont {Claudon}}, \bibinfo {author} {\bibfnamefont {J.-M.}\
  \bibnamefont {G{\'{e}}rard}}, \bibinfo {author} {\bibfnamefont
  {A.}~\bibnamefont {Auff{\`{e}}ves}}, \bibinfo {author} {\bibfnamefont
  {G.}~\bibnamefont {Nogues}}, \bibinfo {author} {\bibfnamefont
  {S.}~\bibnamefont {Seidelin}}, \bibinfo {author} {\bibfnamefont {J.-P.}\
  \bibnamefont {Poizat}}, \bibinfo {author} {\bibfnamefont {O.}~\bibnamefont
  {Arcizet}},\ and\ \bibinfo {author} {\bibfnamefont {M.}~\bibnamefont
  {Richard}},\ }\bibfield  {title} {\bibinfo {title} {{Strain-mediated coupling
  in a quantum dot–mechanical oscillator hybrid system}},\ }\href
  {https://doi.org/10.1038/nnano.2013.274} {\bibfield  {journal} {\bibinfo
  {journal} {Nature Nanotechnology}\ }\textbf {\bibinfo {volume} {9}},\
  \bibinfo {pages} {106} (\bibinfo {year} {2014})}\BibitemShut {NoStop}%
\bibitem [{\citenamefont {Mitchell}\ \emph {et~al.}(2019)\citenamefont
  {Mitchell}, \citenamefont {Lake},\ and\ \citenamefont
  {Barclay}}]{Mitchell-2019-APLphotonics-DiamondMicrodisks}%
  \BibitemOpen
  \bibfield  {author} {\bibinfo {author} {\bibfnamefont {M.}~\bibnamefont
  {Mitchell}}, \bibinfo {author} {\bibfnamefont {D.~P.}\ \bibnamefont {Lake}},\
  and\ \bibinfo {author} {\bibfnamefont {P.~E.}\ \bibnamefont {Barclay}},\
  }\bibfield  {title} {\bibinfo {title} {{Realizing Q {\textgreater} 300 000 in
  diamond microdisks for optomechanics via etch optimization}},\ }\href
  {https://doi.org/10.1063/1.5053122} {\bibfield  {journal} {\bibinfo
  {journal} {APL Photonics}\ }\textbf {\bibinfo {volume} {4}},\ \bibinfo
  {pages} {16101} (\bibinfo {year} {2019})}\BibitemShut {NoStop}%
\bibitem [{\citenamefont {Rokhsari}\ \emph {et~al.}(2005)\citenamefont
  {Rokhsari}, \citenamefont {Kippenberg}, \citenamefont {Carmon},\ and\
  \citenamefont {Vahala}}]{Rokhsari-2005-OpticsExpress-SelfOscillations}%
  \BibitemOpen
  \bibfield  {author} {\bibinfo {author} {\bibfnamefont {H.}~\bibnamefont
  {Rokhsari}}, \bibinfo {author} {\bibfnamefont {T.~J.}\ \bibnamefont
  {Kippenberg}}, \bibinfo {author} {\bibfnamefont {T.}~\bibnamefont {Carmon}},\
  and\ \bibinfo {author} {\bibfnamefont {K.~J.}\ \bibnamefont {Vahala}},\
  }\bibfield  {title} {\bibinfo {title} {{Radiation-pressure-driven
  micro-mechanical oscillator}},\ }\href
  {https://doi.org/10.1364/OPEX.13.005293} {\bibfield  {journal} {\bibinfo
  {journal} {Optics Express}\ }\textbf {\bibinfo {volume} {13}},\ \bibinfo
  {pages} {5293} (\bibinfo {year} {2005})}\BibitemShut {NoStop}%
\bibitem [{\citenamefont {Hossein-Zadeh}\ and\ \citenamefont
  {Vahala}(2008)}]{Hossein-Zadeh-2008-APL-InjectionLockingOptomechanics}%
  \BibitemOpen
  \bibfield  {author} {\bibinfo {author} {\bibfnamefont {M.}~\bibnamefont
  {Hossein-Zadeh}}\ and\ \bibinfo {author} {\bibfnamefont {K.~J.}\ \bibnamefont
  {Vahala}},\ }\bibfield  {title} {\bibinfo {title} {{Observation of injection
  locking in an optomechanical rf oscillator}},\ }\href
  {https://doi.org/10.1063/1.3028024} {\bibfield  {journal} {\bibinfo
  {journal} {Applied Physics Letters}\ }\textbf {\bibinfo {volume} {93}},\
  \bibinfo {pages} {191115} (\bibinfo {year} {2008})}\BibitemShut {NoStop}%
\bibitem [{\citenamefont {Hong}\ and\ \citenamefont
  {Hajimiri}(2019)}]{Hong-InjectionLocking-IEEE-2019}%
  \BibitemOpen
  \bibfield  {author} {\bibinfo {author} {\bibfnamefont {B.}~\bibnamefont
  {Hong}}\ and\ \bibinfo {author} {\bibfnamefont {A.}~\bibnamefont
  {Hajimiri}},\ }\bibfield  {title} {\bibinfo {title} {{A General Theory of
  Injection Locking and Pulling in Electrical Oscillators—Part II: Amplitude
  Modulation in {\$}LC{\$} Oscillators, Transient Behavior, and Frequency
  Division}},\ }\href {https://doi.org/10.1109/JSSC.2019.2908763} {\bibfield
  {journal} {\bibinfo  {journal} {IEEE Journal of Solid-State Circuits}\
  }\textbf {\bibinfo {volume} {54}},\ \bibinfo {pages} {2122} (\bibinfo {year}
  {2019})}\BibitemShut {NoStop}%
\bibitem [{\citenamefont {Gruber}\ \emph {et~al.}(1997)\citenamefont {Gruber},
  \citenamefont {Dr{\"{a}}benstedt}, \citenamefont {Tietz}, \citenamefont
  {Fleury}, \citenamefont {Wrachtrup},\ and\ \citenamefont {{Von
  Borczyskowski}}}]{Gruber-Science-1997-SingleNVdetection}%
  \BibitemOpen
  \bibfield  {author} {\bibinfo {author} {\bibfnamefont {A.}~\bibnamefont
  {Gruber}}, \bibinfo {author} {\bibfnamefont {A.}~\bibnamefont
  {Dr{\"{a}}benstedt}}, \bibinfo {author} {\bibfnamefont {C.}~\bibnamefont
  {Tietz}}, \bibinfo {author} {\bibfnamefont {L.}~\bibnamefont {Fleury}},
  \bibinfo {author} {\bibfnamefont {J.}~\bibnamefont {Wrachtrup}},\ and\
  \bibinfo {author} {\bibfnamefont {C.}~\bibnamefont {{Von Borczyskowski}}},\
  }\bibfield  {title} {\bibinfo {title} {{Scanning confocal optical microscopy
  and magnetic resonance on single defect centers}},\ }\href
  {https://doi.org/10.1126/science.276.5321.2012} {\bibfield  {journal}
  {\bibinfo  {journal} {Science}\ }\textbf {\bibinfo {volume} {276}},\ \bibinfo
  {pages} {2012} (\bibinfo {year} {1997})}\BibitemShut {NoStop}%
\bibitem [{\citenamefont {Udvarhelyi}\ \emph {et~al.}(2018)\citenamefont
  {Udvarhelyi}, \citenamefont {Shkolnikov}, \citenamefont {Gali}, \citenamefont
  {Burkard},\ and\ \citenamefont
  {P{\'{a}}lyi}}]{Udvarhelyi-PRB-2018-NVspinStrainInteractions}%
  \BibitemOpen
  \bibfield  {author} {\bibinfo {author} {\bibfnamefont {P.}~\bibnamefont
  {Udvarhelyi}}, \bibinfo {author} {\bibfnamefont {V.~O.}\ \bibnamefont
  {Shkolnikov}}, \bibinfo {author} {\bibfnamefont {A.}~\bibnamefont {Gali}},
  \bibinfo {author} {\bibfnamefont {G.}~\bibnamefont {Burkard}},\ and\ \bibinfo
  {author} {\bibfnamefont {A.}~\bibnamefont {P{\'{a}}lyi}},\ }\bibfield
  {title} {\bibinfo {title} {{Spin-strain interaction in nitrogen-vacancy
  centers in diamond}},\ }\href {https://doi.org/10.1103/PhysRevB.98.075201}
  {\bibfield  {journal} {\bibinfo  {journal} {Physical Review B}\ }\textbf
  {\bibinfo {volume} {98}},\ \bibinfo {pages} {75201} (\bibinfo {year}
  {2018})}\BibitemShut {NoStop}%
\bibitem [{\citenamefont {MacQuarrie}\ \emph
  {et~al.}(2015{\natexlab{a}})\citenamefont {MacQuarrie}, \citenamefont
  {Gosavi}, \citenamefont {Moehle}, \citenamefont {Jungwirth}, \citenamefont
  {Bhave},\ and\ \citenamefont
  {Fuchs}}]{MacQuarrie-2015-Optica-CoherentControlNVstrain}%
  \BibitemOpen
  \bibfield  {author} {\bibinfo {author} {\bibfnamefont {E.~R.}\ \bibnamefont
  {MacQuarrie}}, \bibinfo {author} {\bibfnamefont {T.~A.}\ \bibnamefont
  {Gosavi}}, \bibinfo {author} {\bibfnamefont {A.~M.}\ \bibnamefont {Moehle}},
  \bibinfo {author} {\bibfnamefont {N.~R.}\ \bibnamefont {Jungwirth}}, \bibinfo
  {author} {\bibfnamefont {S.~A.}\ \bibnamefont {Bhave}},\ and\ \bibinfo
  {author} {\bibfnamefont {G.~D.}\ \bibnamefont {Fuchs}},\ }\bibfield  {title}
  {\bibinfo {title} {{Coherent control of a nitrogen-vacancy center spin
  ensemble with a diamond mechanical resonator}},\ }\href
  {https://doi.org/10.1364/OPTICA.2.000233} {\bibfield  {journal} {\bibinfo
  {journal} {Optica}\ }\textbf {\bibinfo {volume} {2}},\ \bibinfo {pages} {233}
  (\bibinfo {year} {2015}{\natexlab{a}})}\BibitemShut {NoStop}%
\bibitem [{\citenamefont {Hong}\ \emph {et~al.}(2012)\citenamefont {Hong},
  \citenamefont {Grinolds}, \citenamefont {Maletinsky}, \citenamefont
  {Walsworth}, \citenamefont {Lukin},\ and\ \citenamefont
  {Yacoby}}]{Hong-2012-NanoLetters-NVcoupledCanteliver}%
  \BibitemOpen
  \bibfield  {author} {\bibinfo {author} {\bibfnamefont {S.}~\bibnamefont
  {Hong}}, \bibinfo {author} {\bibfnamefont {M.~S.}\ \bibnamefont {Grinolds}},
  \bibinfo {author} {\bibfnamefont {P.}~\bibnamefont {Maletinsky}}, \bibinfo
  {author} {\bibfnamefont {R.~L.}\ \bibnamefont {Walsworth}}, \bibinfo {author}
  {\bibfnamefont {M.~D.}\ \bibnamefont {Lukin}},\ and\ \bibinfo {author}
  {\bibfnamefont {A.}~\bibnamefont {Yacoby}},\ }\bibfield  {title} {\bibinfo
  {title} {{Coherent, Mechanical Control of a Single Electronic Spin}},\ }\href
  {https://doi.org/10.1021/nl300775c} {\bibfield  {journal} {\bibinfo
  {journal} {Nano Letters}\ }\textbf {\bibinfo {volume} {12}},\ \bibinfo
  {pages} {3920} (\bibinfo {year} {2012})}\BibitemShut {NoStop}%
\bibitem [{\citenamefont {MacQuarrie}\ \emph
  {et~al.}(2015{\natexlab{b}})\citenamefont {MacQuarrie}, \citenamefont
  {Gosavi}, \citenamefont {Bhave},\ and\ \citenamefont
  {Fuchs}}]{MacQuarrie-2015-PRB-DynamicalDecouplingSpin}%
  \BibitemOpen
  \bibfield  {author} {\bibinfo {author} {\bibfnamefont {E.~R.}\ \bibnamefont
  {MacQuarrie}}, \bibinfo {author} {\bibfnamefont {T.~A.}\ \bibnamefont
  {Gosavi}}, \bibinfo {author} {\bibfnamefont {S.~A.}\ \bibnamefont {Bhave}},\
  and\ \bibinfo {author} {\bibfnamefont {G.~D.}\ \bibnamefont {Fuchs}},\
  }\bibfield  {title} {\bibinfo {title} {{Continuous dynamical decoupling of a
  single diamond nitrogen-vacancy center spin with a mechanical resonator}},\
  }\href {https://doi.org/10.1103/PhysRevB.92.224419} {\bibfield  {journal}
  {\bibinfo  {journal} {Physical Review B}\ }\textbf {\bibinfo {volume} {92}},\
  \bibinfo {pages} {224419} (\bibinfo {year} {2015}{\natexlab{b}})}\BibitemShut
  {NoStop}%
\bibitem [{\citenamefont {Chen}\ \emph {et~al.}(2018)\citenamefont {Chen},
  \citenamefont {MacQuarrie},\ and\ \citenamefont
  {Fuchs}}]{MacQuarrie-2018-PRL-NVorbitalStrainManipulation}%
  \BibitemOpen
  \bibfield  {author} {\bibinfo {author} {\bibfnamefont {H.~Y.}\ \bibnamefont
  {Chen}}, \bibinfo {author} {\bibfnamefont {E.~R.}\ \bibnamefont
  {MacQuarrie}},\ and\ \bibinfo {author} {\bibfnamefont {G.~D.}\ \bibnamefont
  {Fuchs}},\ }\bibfield  {title} {\bibinfo {title} {{Orbital State Manipulation
  of a Diamond Nitrogen-Vacancy Center Using a Mechanical Resonator}},\ }\href
  {https://doi.org/10.1103/PhysRevLett.120.167401} {\bibfield  {journal}
  {\bibinfo  {journal} {Physical Review Letters}\ }\textbf {\bibinfo {volume}
  {120}},\ \bibinfo {pages} {167401} (\bibinfo {year} {2018})}\BibitemShut
  {NoStop}%
\bibitem [{\citenamefont {Lee}\ \emph {et~al.}(2016)\citenamefont {Lee},
  \citenamefont {Lee}, \citenamefont {Ovartchaiyapong}, \citenamefont
  {Minguzzi}, \citenamefont {Maze},\ and\ \citenamefont {{Bleszynski
  Jayich}}}]{Lee-NVcanteliverTuning-PRApplied-2016}%
  \BibitemOpen
  \bibfield  {author} {\bibinfo {author} {\bibfnamefont {K.~W.}\ \bibnamefont
  {Lee}}, \bibinfo {author} {\bibfnamefont {D.}~\bibnamefont {Lee}}, \bibinfo
  {author} {\bibfnamefont {P.}~\bibnamefont {Ovartchaiyapong}}, \bibinfo
  {author} {\bibfnamefont {J.}~\bibnamefont {Minguzzi}}, \bibinfo {author}
  {\bibfnamefont {J.~R.}\ \bibnamefont {Maze}},\ and\ \bibinfo {author}
  {\bibfnamefont {A.~C.}\ \bibnamefont {{Bleszynski Jayich}}},\ }\bibfield
  {title} {\bibinfo {title} {{Strain Coupling of a Mechanical Resonator to a
  Single Quantum Emitter in Diamond}},\ }\href
  {https://doi.org/10.1103/PhysRevApplied.6.034005} {\bibfield  {journal}
  {\bibinfo  {journal} {Physical Review Applied}\ }\textbf {\bibinfo {volume}
  {6}},\ \bibinfo {pages} {34005} (\bibinfo {year} {2016})}\BibitemShut
  {NoStop}%
\bibitem [{\citenamefont {Poot}\ \emph {et~al.}(2012)\citenamefont {Poot},
  \citenamefont {Fong}, \citenamefont {Bagheri}, \citenamefont {Pernice},\ and\
  \citenamefont {Tang}}]{Poot-PRA-BackactionLimitSO-2012}%
  \BibitemOpen
  \bibfield  {author} {\bibinfo {author} {\bibfnamefont {M.}~\bibnamefont
  {Poot}}, \bibinfo {author} {\bibfnamefont {K.~Y.}\ \bibnamefont {Fong}},
  \bibinfo {author} {\bibfnamefont {M.}~\bibnamefont {Bagheri}}, \bibinfo
  {author} {\bibfnamefont {W.~H.~P.}\ \bibnamefont {Pernice}},\ and\ \bibinfo
  {author} {\bibfnamefont {H.~X.}\ \bibnamefont {Tang}},\ }\bibfield  {title}
  {\bibinfo {title} {{Backaction limits on self-sustained optomechanical
  oscillations}},\ }\href {https://doi.org/10.1103/PhysRevA.86.053826}
  {\bibfield  {journal} {\bibinfo  {journal} {Phys. Rev. A}\ }\textbf {\bibinfo
  {volume} {86}},\ \bibinfo {pages} {53826} (\bibinfo {year}
  {2012})}\BibitemShut {NoStop}%
\bibitem [{\citenamefont {Lake}\ \emph {et~al.}(2021)\citenamefont {Lake},
  \citenamefont {Mitchell}, \citenamefont {Sukachev},\ and\ \citenamefont
  {Barclay}}]{lake2021processing}%
  \BibitemOpen
  \bibfield  {author} {\bibinfo {author} {\bibfnamefont {D.~P.}\ \bibnamefont
  {Lake}}, \bibinfo {author} {\bibfnamefont {M.}~\bibnamefont {Mitchell}},
  \bibinfo {author} {\bibfnamefont {D.~D.}\ \bibnamefont {Sukachev}},\ and\
  \bibinfo {author} {\bibfnamefont {P.~E.}\ \bibnamefont {Barclay}},\
  }\bibfield  {title} {\bibinfo {title} {Processing light with an optically
  tunable mechanical memory},\ }\href@noop {} {\bibfield  {journal} {\bibinfo
  {journal} {Nature communications}\ }\textbf {\bibinfo {volume} {12}},\
  \bibinfo {pages} {1} (\bibinfo {year} {2021})}\BibitemShut {NoStop}%
\bibitem [{\citenamefont {Neuman}\ \emph {et~al.}(2020)\citenamefont {Neuman},
  \citenamefont {Eichenfield}, \citenamefont {Trusheim}, \citenamefont
  {Hackett}, \citenamefont {Narang},\ and\ \citenamefont
  {Englund}}]{neuman-PhotonicBus-2020-arxiv}%
  \BibitemOpen
  \bibfield  {author} {\bibinfo {author} {\bibfnamefont {T.}~\bibnamefont
  {Neuman}}, \bibinfo {author} {\bibfnamefont {M.}~\bibnamefont {Eichenfield}},
  \bibinfo {author} {\bibfnamefont {M.}~\bibnamefont {Trusheim}}, \bibinfo
  {author} {\bibfnamefont {L.}~\bibnamefont {Hackett}}, \bibinfo {author}
  {\bibfnamefont {P.}~\bibnamefont {Narang}},\ and\ \bibinfo {author}
  {\bibfnamefont {D.}~\bibnamefont {Englund}},\ }\href@noop {} {\bibinfo
  {title} {{A Phononic Bus for Coherent Interfaces Between a Superconducting
  Quantum Processor, Spin Memory, and Photonic Quantum Networks}}} (\bibinfo
  {year} {2020}),\ \Eprint {https://arxiv.org/abs/2003.08383} {arXiv:2003.08383
  [quant-ph]} \BibitemShut {NoStop}%
\bibitem [{\citenamefont {Meesala}\ \emph {et~al.}(2018)\citenamefont
  {Meesala}, \citenamefont {Sohn}, \citenamefont {Pingault}, \citenamefont
  {Shao}, \citenamefont {Atikian}, \citenamefont {Holzgrafe}, \citenamefont
  {Gundogan}, \citenamefont {Stavrakas}, \citenamefont {Sipahigil},
  \citenamefont {Chia}, \citenamefont {Burek}, \citenamefont {Zhang},
  \citenamefont {Wu}, \citenamefont {Pacheco}, \citenamefont {Abraham},
  \citenamefont {Bielejec}, \citenamefont {Lukin}, \citenamefont {Atature},\
  and\ \citenamefont {Loncar}}]{Meesala2018}%
  \BibitemOpen
  \bibfield  {author} {\bibinfo {author} {\bibfnamefont {S.}~\bibnamefont
  {Meesala}}, \bibinfo {author} {\bibfnamefont {Y.-I.}\ \bibnamefont {Sohn}},
  \bibinfo {author} {\bibfnamefont {B.}~\bibnamefont {Pingault}}, \bibinfo
  {author} {\bibfnamefont {L.}~\bibnamefont {Shao}}, \bibinfo {author}
  {\bibfnamefont {H.~A.}\ \bibnamefont {Atikian}}, \bibinfo {author}
  {\bibfnamefont {J.}~\bibnamefont {Holzgrafe}}, \bibinfo {author}
  {\bibfnamefont {M.}~\bibnamefont {Gundogan}}, \bibinfo {author}
  {\bibfnamefont {C.}~\bibnamefont {Stavrakas}}, \bibinfo {author}
  {\bibfnamefont {A.}~\bibnamefont {Sipahigil}}, \bibinfo {author}
  {\bibfnamefont {C.}~\bibnamefont {Chia}}, \bibinfo {author} {\bibfnamefont
  {M.~J.}\ \bibnamefont {Burek}}, \bibinfo {author} {\bibfnamefont
  {M.}~\bibnamefont {Zhang}}, \bibinfo {author} {\bibfnamefont
  {L.}~\bibnamefont {Wu}}, \bibinfo {author} {\bibfnamefont {J.~L.}\
  \bibnamefont {Pacheco}}, \bibinfo {author} {\bibfnamefont {J.}~\bibnamefont
  {Abraham}}, \bibinfo {author} {\bibfnamefont {E.}~\bibnamefont {Bielejec}},
  \bibinfo {author} {\bibfnamefont {M.~D.}\ \bibnamefont {Lukin}}, \bibinfo
  {author} {\bibfnamefont {M.}~\bibnamefont {Atature}},\ and\ \bibinfo {author}
  {\bibfnamefont {M.}~\bibnamefont {Loncar}},\ }\bibfield  {title} {\bibinfo
  {title} {{Strain engineering of the silicon-vacancy center in diamond}},\
  }\href {http://arxiv.org/abs/1801.09833} {\  (\bibinfo {year} {2018})},\
  \Eprint {https://arxiv.org/abs/1801.09833} {arXiv:1801.09833} \BibitemShut
  {NoStop}%
\bibitem [{\citenamefont {MacCabe}\ \emph {et~al.}(2020)\citenamefont
  {MacCabe}, \citenamefont {Ren}, \citenamefont {Luo}, \citenamefont {Cohen},
  \citenamefont {Zhou}, \citenamefont {Sipahigil}, \citenamefont
  {Mirhosseini},\ and\ \citenamefont
  {Painter}}]{MacCabe-UltraLongPhononicLifetime-Science-2020}%
  \BibitemOpen
  \bibfield  {author} {\bibinfo {author} {\bibfnamefont {G.~S.}\ \bibnamefont
  {MacCabe}}, \bibinfo {author} {\bibfnamefont {H.}~\bibnamefont {Ren}},
  \bibinfo {author} {\bibfnamefont {J.}~\bibnamefont {Luo}}, \bibinfo {author}
  {\bibfnamefont {J.~D.}\ \bibnamefont {Cohen}}, \bibinfo {author}
  {\bibfnamefont {H.}~\bibnamefont {Zhou}}, \bibinfo {author} {\bibfnamefont
  {A.}~\bibnamefont {Sipahigil}}, \bibinfo {author} {\bibfnamefont
  {M.}~\bibnamefont {Mirhosseini}},\ and\ \bibinfo {author} {\bibfnamefont
  {O.}~\bibnamefont {Painter}},\ }\bibfield  {title} {\bibinfo {title}
  {{Nano-acoustic resonator with ultralong phonon lifetime}},\ }\href
  {https://doi.org/10.1126/science.abc7312} {\bibfield  {journal} {\bibinfo
  {journal} {Science}\ }\textbf {\bibinfo {volume} {370}},\ \bibinfo {pages}
  {840} (\bibinfo {year} {2020})}\BibitemShut {NoStop}%
\bibitem [{\citenamefont {Li}\ \emph {et~al.}(2020)\citenamefont {Li},
  \citenamefont {Zhou}, \citenamefont {Gao},\ and\ \citenamefont
  {Nori}}]{li2020enhancing}%
  \BibitemOpen
  \bibfield  {author} {\bibinfo {author} {\bibfnamefont {P.-B.}\ \bibnamefont
  {Li}}, \bibinfo {author} {\bibfnamefont {Y.}~\bibnamefont {Zhou}}, \bibinfo
  {author} {\bibfnamefont {W.-B.}\ \bibnamefont {Gao}},\ and\ \bibinfo {author}
  {\bibfnamefont {F.}~\bibnamefont {Nori}},\ }\bibfield  {title} {\bibinfo
  {title} {Enhancing spin-phonon and spin-spin interactions using linear
  resources in a hybrid quantum system},\ }\href@noop {} {\bibfield  {journal}
  {\bibinfo  {journal} {Physical Review Letters}\ }\textbf {\bibinfo {volume}
  {125}},\ \bibinfo {pages} {153602} (\bibinfo {year} {2020})}\BibitemShut
  {NoStop}%
\bibitem [{\citenamefont {Chamberland}\ \emph {et~al.}(2020)\citenamefont
  {Chamberland}, \citenamefont {Noh}, \citenamefont {Arrangoiz-Arriola},
  \citenamefont {Campbell}, \citenamefont {Hann}, \citenamefont {Iverson},
  \citenamefont {Putterman}, \citenamefont {Bohdanowicz}, \citenamefont
  {Flammia}, \citenamefont {Keller}, \citenamefont {Refael}, \citenamefont
  {Preskill}, \citenamefont {Jiang}, \citenamefont {Safavi-Naeini},
  \citenamefont {Painter},\ and\ \citenamefont
  {Brand{\~{a}}o}}]{Chamberland--CatStatesMW--Mechanics-2020-arxiv}%
  \BibitemOpen
  \bibfield  {author} {\bibinfo {author} {\bibfnamefont {C.}~\bibnamefont
  {Chamberland}}, \bibinfo {author} {\bibfnamefont {K.}~\bibnamefont {Noh}},
  \bibinfo {author} {\bibfnamefont {P.}~\bibnamefont {Arrangoiz-Arriola}},
  \bibinfo {author} {\bibfnamefont {E.~T.}\ \bibnamefont {Campbell}}, \bibinfo
  {author} {\bibfnamefont {C.~T.}\ \bibnamefont {Hann}}, \bibinfo {author}
  {\bibfnamefont {J.}~\bibnamefont {Iverson}}, \bibinfo {author} {\bibfnamefont
  {H.}~\bibnamefont {Putterman}}, \bibinfo {author} {\bibfnamefont {T.~C.}\
  \bibnamefont {Bohdanowicz}}, \bibinfo {author} {\bibfnamefont {S.~T.}\
  \bibnamefont {Flammia}}, \bibinfo {author} {\bibfnamefont {A.}~\bibnamefont
  {Keller}}, \bibinfo {author} {\bibfnamefont {G.}~\bibnamefont {Refael}},
  \bibinfo {author} {\bibfnamefont {J.}~\bibnamefont {Preskill}}, \bibinfo
  {author} {\bibfnamefont {L.}~\bibnamefont {Jiang}}, \bibinfo {author}
  {\bibfnamefont {A.~H.}\ \bibnamefont {Safavi-Naeini}}, \bibinfo {author}
  {\bibfnamefont {O.}~\bibnamefont {Painter}},\ and\ \bibinfo {author}
  {\bibfnamefont {F.~G. S.~L.}\ \bibnamefont {Brand{\~{a}}o}},\ }\bibfield
  {title} {\bibinfo {title} {{Building a fault-tolerant quantum computer using
  concatenated cat codes}},\ }\href {http://arxiv.org/abs/2012.04108}
  {\bibfield  {journal} {\bibinfo  {journal} {arXiv}\ ,\ \bibinfo {pages}
  {2012.04108}} (\bibinfo {year} {2020})},\ \Eprint
  {https://arxiv.org/abs/2012.04108} {arXiv:2012.04108} \BibitemShut {NoStop}%
\bibitem [{\citenamefont {MacQuarrie}\ \emph {et~al.}(2017)\citenamefont
  {MacQuarrie}, \citenamefont {Otten}, \citenamefont {Gray},\ and\
  \citenamefont {Fuchs}}]{MacQuarrie--SpinCoolingOptomechanics-2017-NatComm}%
  \BibitemOpen
  \bibfield  {author} {\bibinfo {author} {\bibfnamefont {E.~R.}\ \bibnamefont
  {MacQuarrie}}, \bibinfo {author} {\bibfnamefont {M.}~\bibnamefont {Otten}},
  \bibinfo {author} {\bibfnamefont {S.~K.}\ \bibnamefont {Gray}},\ and\
  \bibinfo {author} {\bibfnamefont {G.~D.}\ \bibnamefont {Fuchs}},\ }\bibfield
  {title} {\bibinfo {title} {{Cooling a mechanical resonator with
  nitrogen-vacancy centres using a room temperature excited state spin–strain
  interaction}},\ }\href {https://doi.org/10.1038/ncomms14358} {\bibfield
  {journal} {\bibinfo  {journal} {Nature Communications}\ }\textbf {\bibinfo
  {volume} {8}},\ \bibinfo {pages} {14358} (\bibinfo {year}
  {2017})}\BibitemShut {NoStop}%
\bibitem [{\citenamefont {Kettler}\ \emph {et~al.}(2020)\citenamefont
  {Kettler}, \citenamefont {Vaish}, \citenamefont {de~L{\'{e}}pinay},
  \citenamefont {Besga}, \citenamefont {de~Assis}, \citenamefont {Bourgeois},
  \citenamefont {Auff{\`{e}}ves}, \citenamefont {Richard}, \citenamefont
  {Claudon}, \citenamefont {G{\'{e}}rard}, \citenamefont {Pigeau},
  \citenamefont {Arcizet}, \citenamefont {Verlot},\ and\ \citenamefont
  {Poizat}}]{Kettler-OptomechanicalControlQD-NatNano-2020}%
  \BibitemOpen
  \bibfield  {author} {\bibinfo {author} {\bibfnamefont {J.}~\bibnamefont
  {Kettler}}, \bibinfo {author} {\bibfnamefont {N.}~\bibnamefont {Vaish}},
  \bibinfo {author} {\bibfnamefont {L.~M.}\ \bibnamefont {de~L{\'{e}}pinay}},
  \bibinfo {author} {\bibfnamefont {B.}~\bibnamefont {Besga}}, \bibinfo
  {author} {\bibfnamefont {P.-L.}\ \bibnamefont {de~Assis}}, \bibinfo {author}
  {\bibfnamefont {O.}~\bibnamefont {Bourgeois}}, \bibinfo {author}
  {\bibfnamefont {A.}~\bibnamefont {Auff{\`{e}}ves}}, \bibinfo {author}
  {\bibfnamefont {M.}~\bibnamefont {Richard}}, \bibinfo {author} {\bibfnamefont
  {J.}~\bibnamefont {Claudon}}, \bibinfo {author} {\bibfnamefont {J.-M.}\
  \bibnamefont {G{\'{e}}rard}}, \bibinfo {author} {\bibfnamefont
  {B.}~\bibnamefont {Pigeau}}, \bibinfo {author} {\bibfnamefont
  {O.}~\bibnamefont {Arcizet}}, \bibinfo {author} {\bibfnamefont
  {P.}~\bibnamefont {Verlot}},\ and\ \bibinfo {author} {\bibfnamefont {J.-P.}\
  \bibnamefont {Poizat}},\ }\bibfield  {title} {\bibinfo {title} {{Inducing
  micromechanical motion by optical excitation of a single quantum dot}},\
  }\bibfield  {journal} {\bibinfo  {journal} {Nature Nanotechnology}\ }\href
  {https://doi.org/10.1038/s41565-020-00814-y} {10.1038/s41565-020-00814-y}
  (\bibinfo {year} {2020})\BibitemShut {NoStop}%
\bibitem [{\citenamefont {Ghobadi}\ \emph {et~al.}(2019)\citenamefont
  {Ghobadi}, \citenamefont {Wein}, \citenamefont {Kaviani}, \citenamefont
  {Barclay},\ and\ \citenamefont
  {Simon}}]{Ghobadi-NVoptomechanicalInterfcace-PRA-2019}%
  \BibitemOpen
  \bibfield  {author} {\bibinfo {author} {\bibfnamefont {R.}~\bibnamefont
  {Ghobadi}}, \bibinfo {author} {\bibfnamefont {S.}~\bibnamefont {Wein}},
  \bibinfo {author} {\bibfnamefont {H.}~\bibnamefont {Kaviani}}, \bibinfo
  {author} {\bibfnamefont {P.}~\bibnamefont {Barclay}},\ and\ \bibinfo {author}
  {\bibfnamefont {C.}~\bibnamefont {Simon}},\ }\bibfield  {title} {\bibinfo
  {title} {{Progress toward cryogen-free spin-photon interfaces based on
  nitrogen-vacancy centers and optomechanics}},\ }\href
  {https://doi.org/10.1103/PhysRevA.99.053825} {\bibfield  {journal} {\bibinfo
  {journal} {Phys. Rev. A}\ }\textbf {\bibinfo {volume} {99}},\ \bibinfo
  {pages} {53825} (\bibinfo {year} {2019})}\BibitemShut {NoStop}%
\end{thebibliography}
\end{document}